\documentclass[aps,prx,twocolumn,superscriptaddress,longbibliography]{revtex4-2}
\usepackage[pages=all, color=black, position={current page.south}, placement=bottom, scale=1, opacity=1, vshift=5mm]{background}
\SetBgContents{
	\tt 
}      

\usepackage{amsmath}
\usepackage{amsthm}
\usepackage{amssymb}
\usepackage{times}
\usepackage{braket}
\usepackage{ulem}

\usepackage{tikz}
\usetikzlibrary{quantikz2} 

\usepackage{dsfont}

\usepackage[utf8]{inputenc}
\usepackage{hyperref}

\usepackage{graphicx, color}
\graphicspath{{fig/}}

\newcommand{\tr}[1]{\mathrm{tr}\left ( #1 \right )}

\usepackage{algorithm, algpseudocode} 
\usepackage{mathrsfs} 
\usepackage{lipsum}

\begin{document}

\title{Mitigating quantum operation infidelity through engineering the distribution of photon losses}

\author{F.~H.~B.~Somhorst}
\affiliation{MESA+ Institute for Nanotechnology, University of Twente, P.~O.~box 217, 7500 AE Enschede, The Netherlands} 
\author{J.~J.~Renema}
\email{j.j.renema@utwente.nl}
\affiliation{MESA+ Institute for Nanotechnology, University of Twente, P.~O.~box 217, 7500 AE Enschede, The Netherlands} 

\begin{abstract}
Multiport interferometers can be constructed from two-port components in various configurations. We investigate how these configurations influence the performance of quantum operations through asymmetries in optical losses. Using numerical simulations, we analyze the effect of photon-loss distributions on the fidelity of operations involving measurements. For both full- and partial-measurement protocols, we compare rectangular (symmetric-loss) and triangular (asymmetric-loss) architectures. Our results show that asymmetric loss configurations can reduce operation infidelity in several cases, revealing a quantifiable trade-off between fidelity and success probability, with implications for the design of high-fidelity photonic circuits.
\end{abstract}

\maketitle

\section{Introduction}
\label{sec:intro}
Integrated linear optical circuits form the backbone of scalable photonic quantum information processing technologies \cite{carolan2015universal,adcock2019programmable,llewellyn2020chip,li2022quantum,moody2022,pai2023experimental,huang2024demonstration,hoch2025quantum,li2025realizing,salavrakos2025photon,psiquantum2025manufacturable}. Miniaturized optical components, such as beam splitters and phase shifters, enable the construction of compact, large-scale interferometric networks. These multiport interferometers provide precise control over quantum states of light, and their implementation as photonic integrated circuits (PICs) enhances resilience to environmental noise (e.g., vibrations and temperature fluctuations), which is critical for high-fidelity, coherent quantum state manipulation \cite{miller1969integrated}.

Any arbitrary unitary transformation acting on a set of optical modes can be decomposed into a sequence of unitary operations, each acting on a two-mode subspace, consistent with the capabilities of linear optical components in quantum photonic systems \cite{nielsen2010quantum,reck1994experimental,clements2016optimal}. Two multiport interferometer design strategies arise from this framework: the \textit{triangular} design (Reck \textit{et al.} \cite{reck1994experimental}) and the \textit{rectangular} design (Clements \textit{et al.} \cite{clements2016optimal}), which differ in optical depth and loss asymmetry. Both designs are special cases within a unified framework of interferometer architectures that can be transformed systematically into one another \cite{vandebril2011chasing,clement2022lov}.

\begin{figure}
\centering
\includegraphics[width=8.6cm, height=20cm, keepaspectratio,]{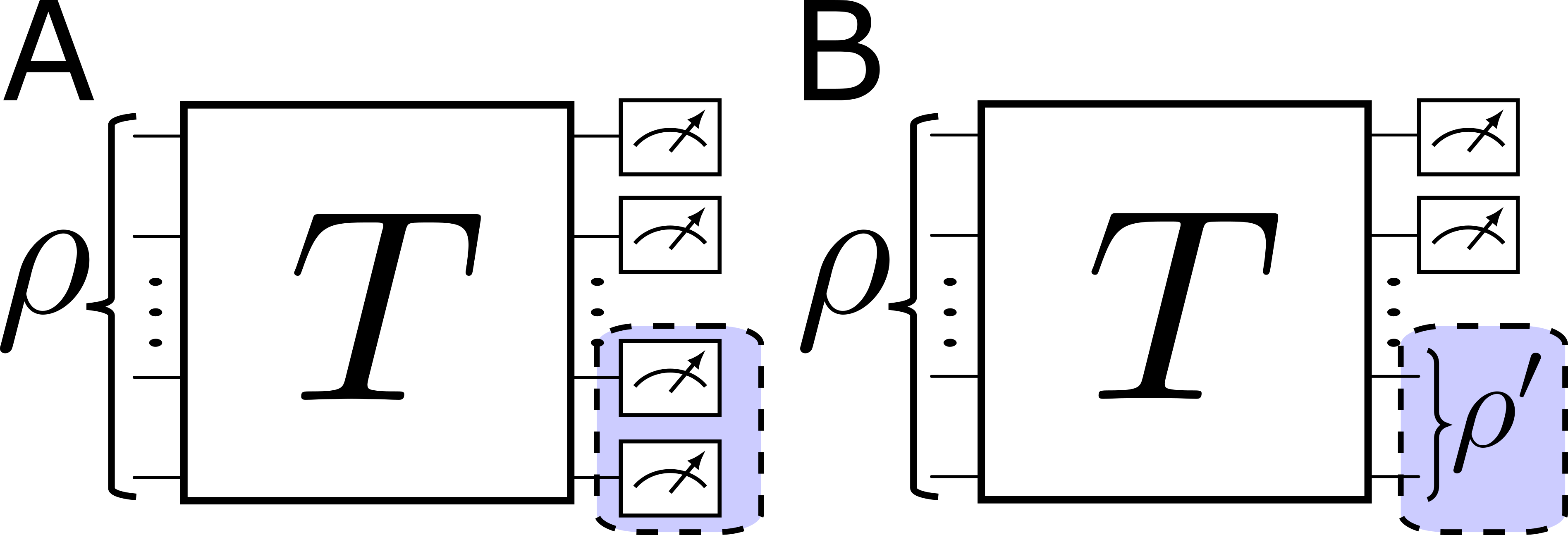}
\caption{\textbf{Classification of linear optical quantum operations.} A quantum channel $\rho \rightarrow \mathcal{T}(\rho)$ is implemented via a multiport interferometer executing a linear transformation $T$ on creation operators, followed by projective measurements. (A) \textit{Full-measurement} operation: all output modes are measured, yielding classical information. (B) \textit{Partial-measurement} operation: one or more output modes are measured. Postselection on specific outcomes produces a heralded state $\rho^\prime$, representing the conditional quantum information output.}
\label{fig:classification}
\end{figure}

Integrated linear optical circuits enable universal photonic quantum information processing when combined with projective measurements on one or more output modes and feedforward operations, which introduce the effective nonlinearity required for entangling gates and conditional operations \cite{knill2001scheme}. This measurement-induced nonlinearity naturally classifies linear optical quantum operations into two categories (Fig. \ref{fig:classification}): \textit{full-measurement} operations, where all output modes are measured, commonly used in tasks such as boson sampling \cite{aaronson2011computational,brod2019photonic}, and \textit{partial-measurement} operations, where only a subset of modes is measured, leaving quantum information in the remaining modes for further processing. In partial-measurement operations, postselection on specific measurement outcomes heralds successful operations, including entangled-state generation \cite{forbes2025heralded} (e.g., Greenberger-Horne-Zeilinger (GHZ) states \cite{gubarev2020improved}) or in photon distillation \cite{marshall2022distillation,somhorst2024photon,saied2024general}.

Circuit design optimization typically targets maximum success probability under ideal conditions \cite{gubarev2020improved}. However, realistic imperfections including photon loss, transformation matrix infidelity, and photonic indistinguishability errors degrade feedforward fidelity \cite{flamini2017benchmarking,shaw2023errors,saied2024advancing}, even with low-loss photonic integrated circuits PICs \cite{chanana2022ultra,taballione20198,taballione202320}. Consequently, maximizing success probability does not necessarily minimize quantum operation loss in practical implementations.

Rectangular multiport interferometers are widely regarded as the most effective design for mitigating photon loss by preserving symmetric interference patterns, producing uniform attenuation upon postselection \cite{clements2016optimal,pai2023experimental,huang2024demonstration,hoch2025quantum,li2025realizing,salavrakos2025photon,zhou2022photonic,erhard2020advances,bai2023photonic,bandyopadhyay2024single,kitayama2019novel}. Symmetric loss also minimizes attenuation for tasks such as Bell state measurements \cite{hilaire2023linear}. By contrast, asymmetric losses can introduce sampling bias, potentially lowering the fidelity of measured quantum information, although certain protocols including Bell inequality violation tests are designed to tolerate or exploit asymmetric detection efficiencies \cite{giustina2013bell,eberhard1993background,brunner2007detection}.

This work numerically examines how photon-loss distributions affect different quantum operations. We define success metrics for both full-measurement and partial-measurement operations and evaluate them across a variety of linear optical tasks. For the examples studied, unbalanced designs outperform balanced designs in full-measurement tasks unless postselection preserves photon number, while in partial-measurement tasks unbalanced designs improve the fidelity of the produced states. These results demonstrate that partial-measurement operations can benefit from asymmetric multiport designs, contrary to the assumption that symmetric configurations are always optimal.

The paper is organized as follows. We first review the framework for linear optical transformations and extend it to include losses. Next we introduce metrics for assessing the efficiency of quantum operations and analyze the effect of postselection in boson sampling. Finally we demonstrate the advantages of asymmetric loss distributions by evaluating rectangular and triangular multiport interferometers in quantum state engineering tasks including photon distillation and 3-qubit GHZ state generation.

\section{Preliminaries}
\label{sec:preliminaries}

\subsection{Linear transformations in lossy systems}
We first introduce unitary transformations in linear quantum optics and extend the framework to include non-unitary transformations. As shown in Fig. \ref{fig:classification}, a linear optical quantum operation $\mathcal{T}(\rho)$ consists of a multiport interferometer and projective measurements\footnote{Here, $\rho$ may represent a state within a larger entangled system, such as in fusion gates \cite{bartolucci2023fusion}.}. The interference component of any quantum operation is fully described by a transformation matrix $T$, which represents a lossy multiport interferometer at the system level.

Without photon losses, the multiport interferometer executes a unitary operation $\mathcal{U}$, which results in the input state being transformed as $\rho \rightarrow \mathcal{U} \rho \mathcal{U}^\dagger$. However, linear optical elements, such as beam splitters and phase shifters, act on the modes of the quantized electromagnetic field, so the unitary operator can be expressed as a transformation of the bosonic creation operators:
\begin{equation}
       \hat{a}_j^\dagger \rightarrow \sum_{i = 1} U_{ij} \hat{a}_i^\dagger,
\end{equation}
where the subscript $i$ ($j$) indicates the output (input) port label, and $U$ is a unitary matrix. 

In the presence of photon losses, the multiport interferometer executes a non-unitary operation expressed by the Kraus operators $E_k$ as $\rho \rightarrow \sum_k E_k\rho E_k^\dagger$. Analogous to the lossless scenario, this non-unitary operation is represented by a transformation matrix $T$ that operates on the creation operators:
\begin{equation}
    \hat{a}_j^\dagger \rightarrow \sum_{i = 1} T_{ij} \hat{a}_i^\dagger,
\end{equation}
where $T^\dagger T \leq \mathcal{I}$.

The sub-unitary matrix $T$ is a submatrix of a larger unitary matrix $\tilde{U}$, with $T$ describing the coupling between measurable (‘\textit{nominal}’) input and output modes \cite{tischler2018quantum}. The purified transformation $\tilde{U}$ additionally accounts for coupling to unmeasurable (‘\textit{virtual}’) modes, representing losses. For any lossy transformation $T$, a corresponding purified transformation $\tilde{U}$ can be constructed using the methods of Ref. \cite{tischler2018quantum}. This construction provides a straightforward framework for analyzing the effects of loss on quantum operations, as including virtual modes allows all photons to be tracked. We use this computational approach throughout this work.

\subsection{Emergence of lossy transformations}

\begin{figure}
    \centering
     \includegraphics[width=8.6cm, height=20cm, keepaspectratio,]{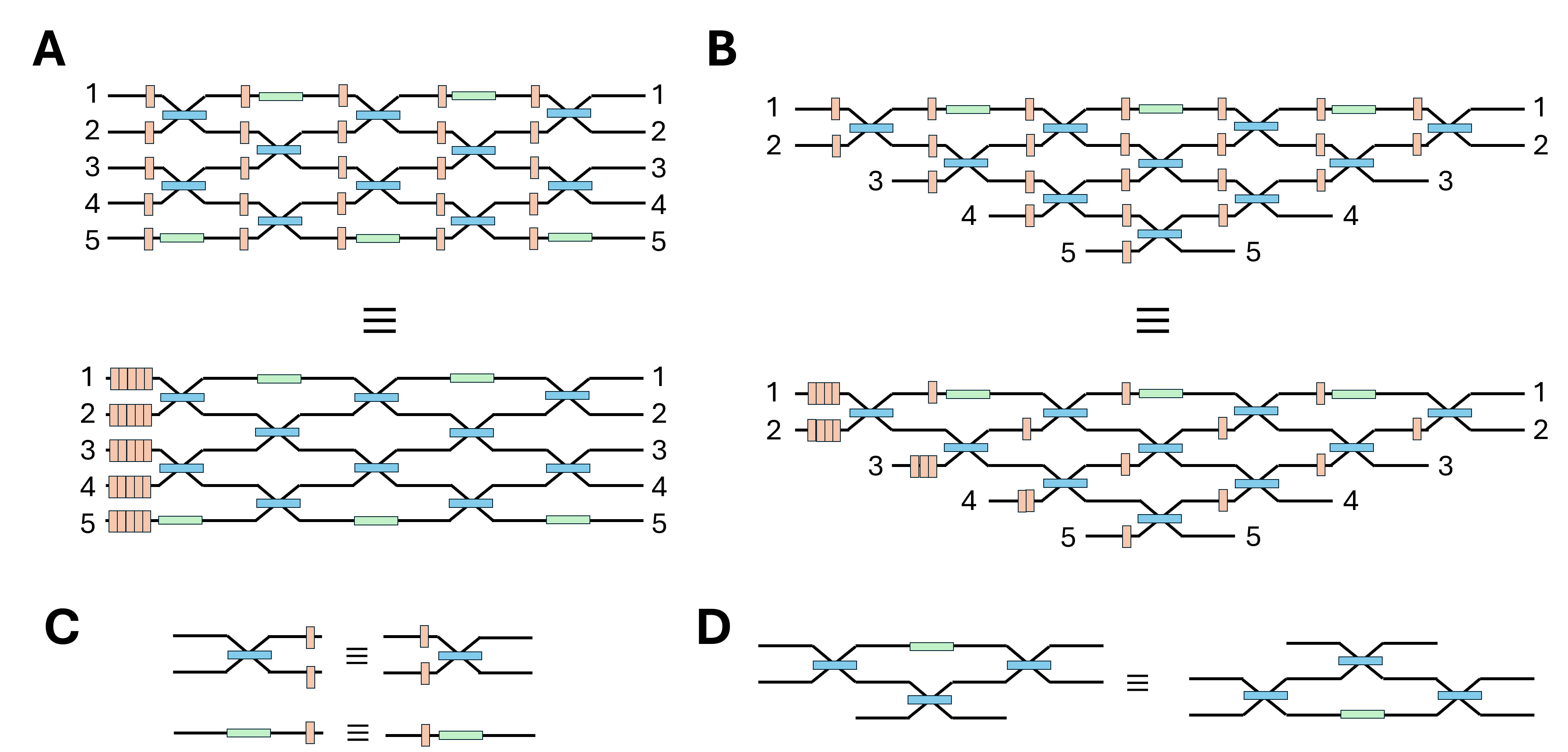}
    \caption{\textbf{Universal 5-port interferometer designs.} Blue rectangles are configurable two-mode unit cells, and green rectangles are static single-mode identity transformations used to equalize path lengths. Orange rectangles represent uniform local loss channels with transmission efficiency $\eta$. A) Rectangular design (top) and its equivalent loss-commuted network (bottom), producing symmetric losses. B) Triangular design (top) and its equivalent loss-commuted network (bottom), producing asymmetric losses; not all loss elements can be commuted to the input. C) Commutation rules for uniform loss elements \cite{oszmaniec2018classical,brod2020classical} used to construct the equivalent networks in A and B. D) Local transformation rule for a subcircuit of two-mode unit cells, enabling conversion between the illustrated designs \cite{clement2022lov}.}
    \label{fig:design}
\end{figure}

The interferometer design determines the structure of $T$. Any multiport interferometer can be constructed from a network of universal two-port interferometers, or \textit{unit cells} (Fig. \ref{fig:design}). Among all universal network configurations, the triangular \cite{reck1994experimental} and rectangular \cite{clements2016optimal} designs are the most widely used, though intermediate designs can be generated using specific transformation rules \cite{vandebril2011chasing,clement2022lov}. While all universal interferometer designs are equivalent in principle, they differ in their handling of photon losses, producing different $T$-matrices. This work focuses on rectangular and triangular designs, which represent the extremes of photon-loss distributions.

We model each unit cell with uniform single-photon transmission efficiency $\eta$ to account for balanced propagation losses \cite{melkozerov2024analysis}. Unit cells are implemented using Mach-Zehnder interferometers (MZIs) with single-mode phase shifters, though more compact implementations exist \cite{bell2021further}. Internal path lengths are matched within the photon coherence length to enable interference \cite{clements2016optimal}, justifying the assumption of uniform propagation loss.

To analyze how $T$ depends on design, we use the lumped-element models in Fig. \ref{fig:design}. Each lossy two-port unit cell is represented as two single-mode loss channels, each with transmission $\eta$, followed by a two-mode unitary transformation. Each loss channel couples to a virtual mode via a beam splitter-like interaction:
\begin{equation}
\hat{a}_j^\dagger \rightarrow \sqrt{\eta} \hat{a}_j^\dagger + \sqrt{1-\eta} \hat{v}_j^\dagger,
\end{equation}
where $\hat{v}_j^\dagger$ is inaccessible. This models amplitude-damping errors with $0 \le \eta \le 1$. Loss channels are assigned to single-port identity transformations, which experience similar on-chip propagation losses. Losses independent of the unit cell network (e.g., source, coupling, or detection losses) are omitted, as this work focuses on photon loss arising from interferometer design.

Fig. \ref{fig:design} shows that linear optical circuits can exhibit significant asymmetric losses even when local components have symmetric losses. Because symmetric losses commute with linear optical transformations \cite{oszmaniec2018classical,brod2020classical}, some loss channels can be shifted to the input ports, providing a clear representation of the overall loss distribution. We represent this distribution with a matrix $\Gamma$, where $|\Gamma_{ij}|^2$ gives the single-photon transmittance from input $j$ to output $i$. The structure of $\Gamma$ depends on the interferometer design. For an $m$-port rectangular design, the matrix elements are defined by:
\begin{equation}
    \Gamma_{ij} = \sqrt{\eta}^m.
    \label{eq:Gamma_rectangular}
\end{equation} 
For a triangular design, however, the matrix elements are defined by:
\begin{eqnarray}
  \Gamma_{ij} := \begin{cases}
    \sqrt{\eta}^{1 + 2m -i -j} &2\leq i,j \leq m,\\
    \Gamma_{i,2} &2 \leq i \leq m,j = 1,\\
    \Gamma_{2,j} &i = 1, 2 \leq j \leq m,\\
    \Gamma_{2,2} & i = 1, j = 1.
  \end{cases}
  \label{eq:Gamma_triangular}
\end{eqnarray}
These definitions show how interferometer design affects $T$. Each $\Gamma_{ij}$ characterizes the amplitude-damping error of the corresponding input-output channel. The lossy $T$-matrix is obtained by element-wise multiplication of $\Gamma$ with a unitary matrix $U$:
\begin{equation}
    T = \Gamma \odot U,
    \label{eq:LossyT}
\end{equation}
where $T_{ij} = \Gamma_{ij} U_{ij}$. The dependence of $\Gamma$ on design directly determines the structure of $T$. In this work, Eq. \ref{eq:LossyT} and both $\Gamma$ definitions are used to compute lossy $T$-matrices for arbitrary unitary matrices $U$.

\subsection{Performance of lossy transformations}
To evaluate and compare different designs implementing partial-measurement quantum operations, we use two figures of merit:
\begin{itemize}
    \item Circuit fidelity
    \item Conditional transmittance
\end{itemize}

Circuit fidelity is often quantified using the normalized average fidelity of single-photon state transformations \cite{clements2016optimal,flamini2017benchmarking}:
\begin{equation}
    \mathcal{F}(U,T) = \left|\frac{\tr{U^\dagger T}}{\sqrt{\tr{U^\dagger U}  \tr{T^\dagger T}}} \right|^2.
    \label{eq:Fidelity_Clements}
\end{equation}

This definition assumes postselection on full transmission of all photons. In many full-measurement schemes, such as boson sampling, this assumption is invalid because full transmission is technologically not achievable \cite{wang2019boson} or the input photon number is indeterminate \cite{zhong2020quantum,zhong2021phase,madsen2022quantum,deng2023gaussian}. In partial-measurement operations, unmeasured output modes are more critical as they carry quantum information and remain unmeasured during operation. We therefore shift our focus from the \textit{input} to the \textit{output} single-photon states. We study the fidelity of states as seen from the viewpoint of the output ports to assess the impact of photon losses on modes sensitive to quantum channel loss. For a single-photon state in output mode $i$,  $\ket{\psi_i} := \hat{a}_i^\dagger \ket{0}$, the transformed state under a matrix $M$ is
\begin{equation}
    \ket{\psi_i(M)} = \sum_{j}^{m}M_{ij}\hat{a}_j^\dagger \ket{0}. 
\end{equation}
The fidelity of a lossy circuit is then
\begin{equation}
    \tilde{\mathcal{F}}_i=|\braket{\psi_i(U)|\psi_i(T)}|^2 = |\sum_{j}^{m}U_{ij}^\dagger T_{ij}|^2,
    \label{eq:Fidelity_Preimage}
\end{equation}
and the average fidelity is
\begin{equation}
    \tilde{\mathcal{F}}_{\text{avg}} := \frac{1}{m} \sum_{i}^{m}|\sum_{j}^{m}U_{ij}^\dagger T_{ij}|^2.
\end{equation}
Because multiphoton transformations depend on $T$, these fidelities provide heuristic insight into the performance of multiphoton operations.

A common figure of merit for partial-measurement operations is the circuit success probability $p_s$, defined by detecting a specific photon pattern in the measured modes. However, $p_s$ does not capture vacuum errors: even when the measurement indicates success, the output state $\rho^\prime$ may be empty. To account for this, we define the conditional transmittance:
\begin{equation}
    \Lambda := \tr{\rho^\prime (\mathcal{I} - \ket{0}\bra{0})}, \quad 0 \le \Lambda \le 1.
    \label{eq:FidelityConditionel}
\end{equation}
Here, $1-\Lambda$ is the probability that the heralded state is empty. Ideally, $\Lambda=1$, which occurs when $T = U$.

\section{Results}
\label{sec:results}

\subsection{Boson Sampling}
Boson sampling is a quantum operation designed to demonstrate an unambiguous quantum computational advantage \cite{aaronson2011computational,brod2019photonic}. It consists of sampling the output distribution of indistinguishable photons after they pass through a multiport interferometer, classifying it as a full-measurement operation. For meaningful boson sampling, the implemented unitary matrices $U$ must be Haar-random \cite{mezzadri2006generate}, making circuit fidelity a primary metric for performance evaluation \cite{clements2016optimal}.

\begin{figure}[h]
\centering
\includegraphics[width=8.6cm, height=20cm, keepaspectratio,]{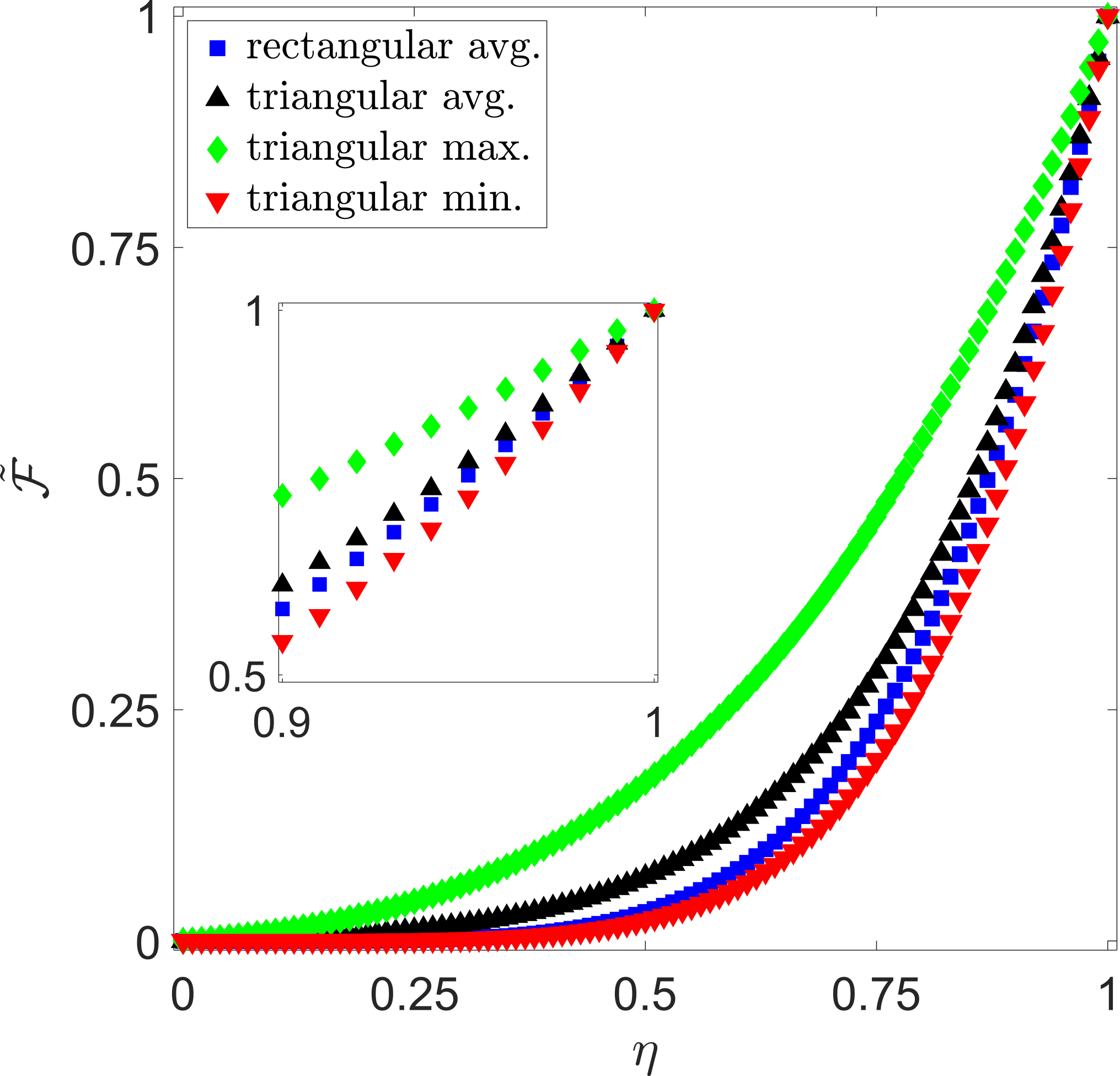}
\caption{\textbf{Single-photon fidelity in 5-mode interferometers under loss.} Non-postselected single-photon fidelity, $\tilde{\mathcal{F}}$, as a function of single-photon transmission efficiency per unit cell, $\eta$. Triangular designs show higher average fidelity than rectangular designs. Rectangular designs maintain uniform fidelity across all output modes, while triangular designs exhibit variation, indicated by the spread between maximum and minimum fidelities. Each point averages over 500 Haar-random unitary samples. Inset: Detailed view of the high-efficiency regime.}
\label{fig:ReimageFidelity}
\end{figure}

We evaluate the non-postselected fidelity (Eq. \ref{eq:Fidelity_Preimage}) to quantify the total impact of photon losses on the operation. Fig. \ref{fig:ReimageFidelity} shows the average fidelity of different 5-mode interferometer designs as a function of single-photon transmission efficiency per unit cell, $\eta$, using Haar-random $U$. The triangular design exhibits higher average fidelity than the rectangular design. Fidelity varies across output modes in the triangular design due to asymmetric losses; we highlight results for modes with minimum fidelity ($i=1,2$) and maximum fidelity ($i=5$). These results indicate that the common preference for rectangular designs arises from postselection bias in Eq. \ref{eq:Fidelity_Clements}.

For partial-measurement quantum operations, directing quantum information to the highest-fidelity output mode reduces susceptibility to photon-loss–induced infidelity. Tracing the photon path to output mode 5 in the triangular design shows that it encounters fewer lossy elements than in the rectangular design. Unlike Ref. \cite{clements2016optimal}, we include propagation losses for single-mode identity transformations in the boundary paths. As discussed in the supplemental document, accounting for these losses slightly improves postselected circuit fidelity relative to the results reported in Ref. \cite{clements2016optimal}.

\subsection{Photon distillation}
Photon distillation is a scalable error suppression protocol that generates highly identical single-photon states, reducing resource costs in photonic quantum computing architectures \cite{somhorst2024photon,saied2024general}. The operation produces a higher-quality photon from multiple lower-quality photons through multiphoton interference, making it a partial-measurement task. While distillation reduces indistinguishability errors, it introduces additional photon loss. Conditional transmittance is therefore the primary metric for evaluating photon-distillation circuit performance.

The relevant unitary transformation $U$ is the $N$-mode Fourier transform, with matrix elements $U_{jk} = \frac{1}{\sqrt{N}} \exp \left( i \frac{2\pi (j-1)(k-1)}{N} \right)$. In photon distillation, a successful operation produces an unentangled state in a single output mode. Asymmetric losses alter the distribution of success probabilities but affect identical and non-identical photons equally, leaving the reduction of indistinguishability errors independent of the design-specific $T$-matrix.

The conditional transmittance for photon-distillation circuits depends on the unit cell single-photon transmission efficiency $\eta$ and the circuit size $N$. For rectangular designs, it scales as
\begin{equation}
\Lambda_{\text{rectangular}} = \eta^N,
\end{equation}
while for triangular designs, it scales as
\begin{equation}
\Lambda_{\text{triangular}} = \frac{N(1-\eta)\eta^N}{2\eta - \eta^2 - \eta^N}.
\end{equation}

These expressions follow from the lossless circuit success probability \cite{somhorst2024photon}:
\begin{equation}
p_0 = \sum_{j=0}^{N-1} (-1)^j (j+1) \prod_{i=1}^{j} \left(1 - \frac{i}{N}\right) + \mathcal{O}(\epsilon),
\end{equation}
where $\epsilon$ quantifies photon indistinguishability errors. The zero-transmission law \cite{tichy2010zero} ensures that incorrect $N$-photon distributions cannot produce valid $N-1$ photon heralds, making photon-distillation operations inherently robust to loss up to $\mathcal{O}(\epsilon)$.

For rectangular designs, all losses can be commuted to the input (Fig. \ref{fig:design}A). The survival probability of the $N-1$ auxiliary photons is $(\eta^N)^{N-1}$, and the survival probability of the distilled photon is $\eta^N$. The circuit success probability is then
\begin{equation}
    p_{\text{s,rectangular}} = p_0 \eta^{N(N-1)} + \mathcal{O}(\epsilon),
    \label{eq:PD_Clements_p_s}
\end{equation}
with conditional transmittance
\begin{equation}
    \Lambda_{\text{rectangular}} = \eta^N + \mathcal{O}(\epsilon).
\end{equation}

For triangular designs, not all losses can be commuted to the input (Fig. \ref{fig:design}B). However, the path to the output mode carrying the distilled photon can remain loss-free. Assuming a balanced $U$, each input mode contributes equally with probability $1/N$. Averaging over all combinations of tracing back the input losses of the distilled photon gives the circuit probability $p_{\text{s,triangular}}$. In other words, this statement implies:  
\begin{equation}
\begin{split}
    p_{\text{s,triangular}} &= \frac{1}{N} \left ( \eta^{-(N-1)} + \sum_{i=1}^{N-1} \eta^{-i} \right )p_{\text{s,triangular}}\Lambda_{\text{triangular}} \\&+ \mathcal{O}(\epsilon),
\end{split}
\end{equation}
which implies (after simplifications):
\begin{equation}
    \Lambda_{\text{triangular}} = \frac{N(1-\eta)\eta^N}{2\eta - \eta^2 - \eta^N} + \mathcal{O}(\epsilon).
    \label{eq:p_erasure_reck}
\end{equation}
Triangular designs therefore yield higher conditional transmittance than rectangular designs for most $\eta < 1$, and this advantage increases with circuit size $N$.

\begin{figure}[h!]
\centering
\includegraphics[width=8.6cm, height=20cm, keepaspectratio,]{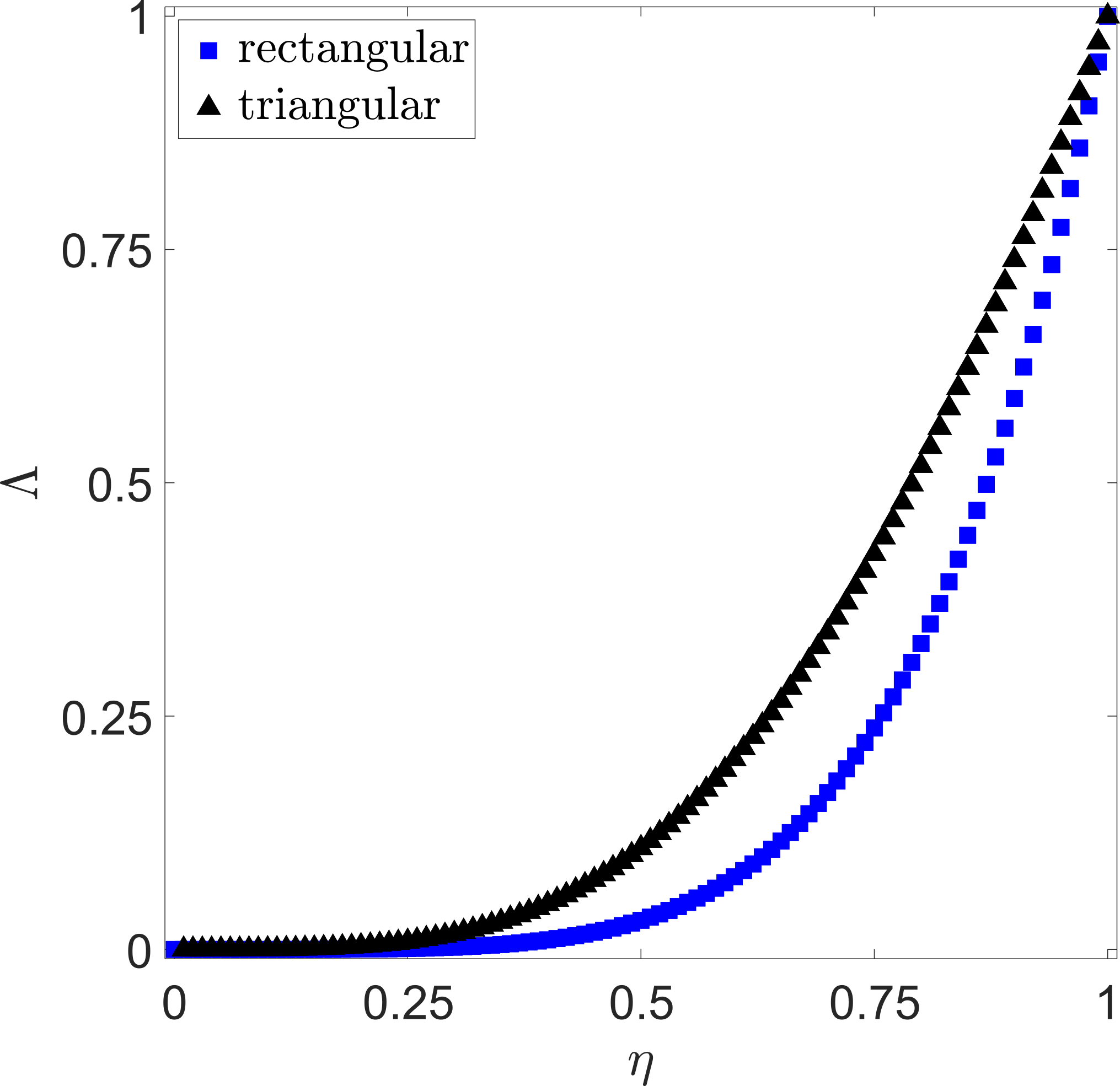}
\caption{\textbf{Conditional transmittance in a 5-photon Fourier transform distillation circuit.} Conditional transmittance $\Lambda$ is plotted as a function of single-photon transmission efficiency $\eta$ for rectangular and triangular designs. Triangular designs yield higher conditional transmittance than rectangular designs for $0 < \eta < 1$.}
\label{fig:PD_N_5_transmittance}
\end{figure}

Fig. \ref{fig:PD_N_5_transmittance} shows the conditional transmittance for a 5-photon Fourier-transform distillation circuit. Across $0 < \eta < 1$, the triangular design achieves higher conditional transmittance than the rectangular design. The advantage of triangular configurations grows with circuit size due to the prefactor $N$ in $\Lambda_{\text{triangular}}$.

\subsection{Greenberger-Horne-Zeilinger states}
GHZ-state generation circuits produce highly entangled quantum states, with the 3-qubit GHZ (GHZ-3) state forming a key building block for scalable and fault-tolerant photonic quantum computing \cite{varnava2008good,bartolucci2023fusion,chen2024heralded}. This operation entangles three photons via multiphoton interference and projective measurements on auxiliary photons, classifying it as a partial-measurement operation \cite{gubarev2020improved}. Minimizing additional losses introduced by the multiport interferometer is critical, therefore, the conditional transmittance of each qubit serves as the primary metric for evaluating GHZ-state generation circuits.

Recent experimental demonstrations of on-chip heralded GHZ-3 generation \cite{chen2024heralded} achieved a total heralding efficiency of $\sim 0.0005$, highlighting scaling limitations. The on-chip unit cell efficiency of the rectangular multiport interferometer used is $\eta = 0.9848$ (see supplemental document), indicating potential improvements via circuit design modifications.

\begin{figure}[h!]
\centering
\includegraphics[width=8.6cm, height=20cm, keepaspectratio,]{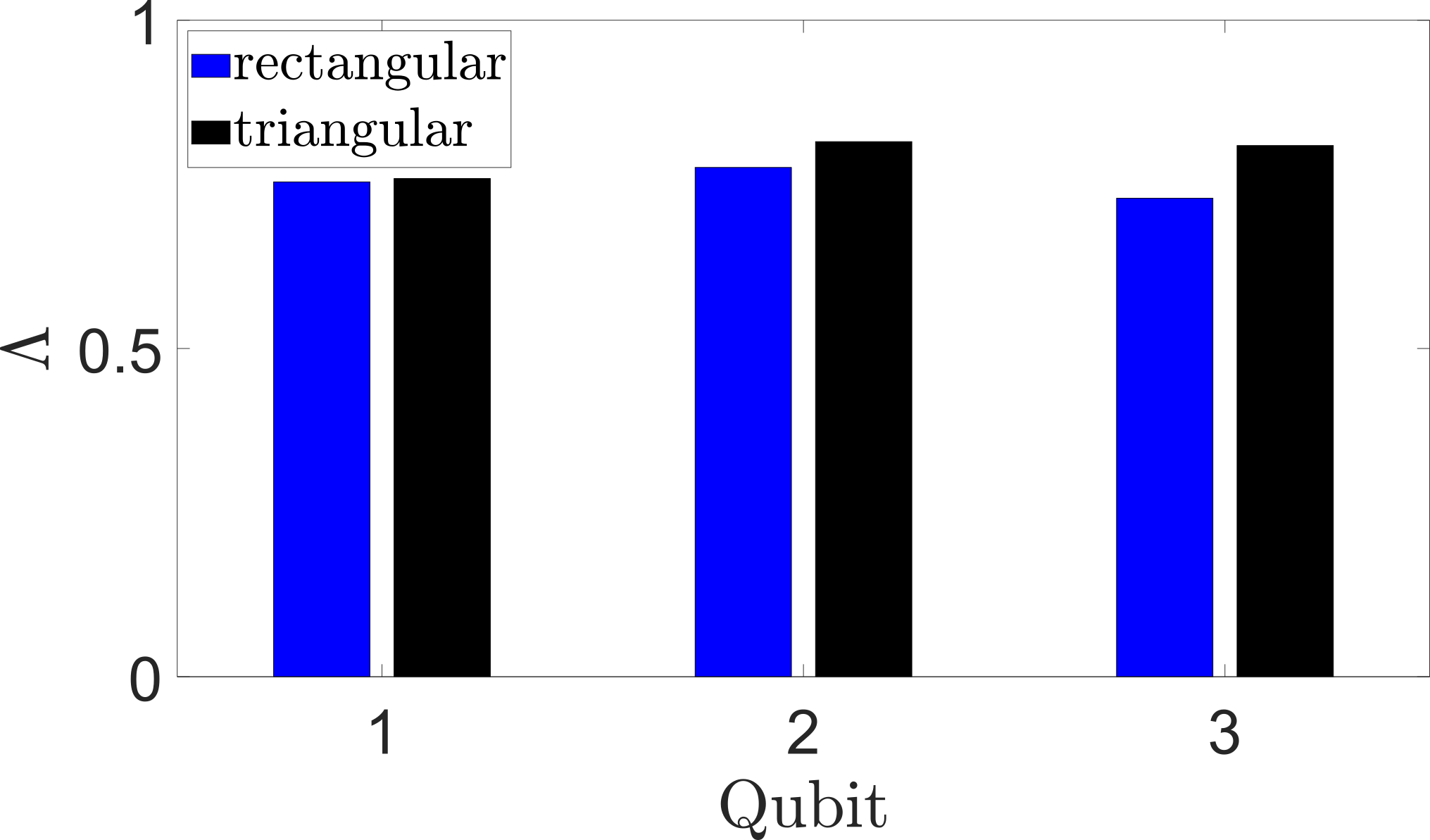}
\caption{\textbf{Conditional transmittance of qubits in heralded 3-qubit GHZ states.} The rectangular design exhibits symmetric losses, yet the marginal conditional qubit transmittance $\Lambda$ is non-uniform. Transitioning to a triangular design reduces this non-uniformity, particularly for qubits 2 and 3. Calculations assume unit cell efficiency $\eta = 0.9848$.}
\label{fig:GHZ3_qubiterasure}
\end{figure}

Fig. \ref{fig:GHZ3_qubiterasure} shows that in the rectangular design, conditional qubit transmittance is non-uniform despite symmetric losses. This non-uniformity is attributed to events involving more than three photons in the measured modes, which non-trivially map onto the postselected three-photon herald measurement due to losses (see supplemental document). Using a triangular multiport interferometer improves the conditional transmittance for all qubits and reduces the disparity between qubits 2 and 3, even though the triangular design introduces asymmetric losses. This improvement is achieved with a slight increase in stabilizer measurement errors (see supplemental document).

\section{Discussion}
In summary, we showed that asymmetric multiport interferometer designs can outperform symmetric designs for partial-measurement quantum operations. Specifically, adopting a triangular rather than a rectangular design reduces vacuum errors in photon-distillation and GHZ-state generation circuits, although it slightly decreases the overall circuit success probability. This reduction in success probability can be mitigated through multiplexing, which introduces a fixed overhead for each operation \cite{migdall2002tailoring}.

Our results provide a framework to evaluate whether balanced or unbalanced loss distributions are optimal for other linear optical circuits. Future research could explore whether intermediate interferometer designs, between triangular and rectangular, offer further improvements, and whether a general criterion can predict the optimal loss distribution without detailed circuit-specific simulations.

For scalable and fault-tolerant linear optical quantum computing, it is critical that photonic circuits generate entangled states with independently degraded qubits, i.e., each qubit experiencing uniform and uncorrelated loss \cite{varnava2008good}. Our findings indicate that symmetric interferometer designs do not necessarily yield uniform qubit losses, highlighting the importance of accounting for photon loss distributions at the circuit design stage. 

\section{Competing interests}
The authors declare that they have no conflict of interest.

\section{Data availability}
All simulated data and analysis scripts generated in this study are available in the 4TU.ResearchData repository (DOI: 10.4121/aac3050c-2cb9-458b-a483-04b875c71cd6).

\section{Acknowledgements}
We thank S.N. van den Hoven, P.W.H. Pinkse, and H. Chrzanowski for scientific discussions. We
thank J. Saied for helpful comments on our first draft of the manuscript. This research is supported by the Photonic Integrated Technology Center (PITC). This publication is part of the project "At the Quantum Edge" of the research programme VIDI, which is financed by the Dutch Research Council (NWO).

\appendix

\section{Impact of single-mode identity transformation losses}
\label{app:LTmultiport}
We evaluate the resilience of multiport interferometers to unbalanced losses following the method of Ref. \cite{clements2016optimal}. For $N = 20$, we generate 500 Haar-random unitary matrices $U$. Uniform losses $l$ [dB] are applied to the input of each two-mode and single-mode transformation, yielding a single-photon transmission efficiency $\eta = 10^{-l/10}$. Using Eqs. 4 and 5, we compute the loss matrices $\Gamma$ for rectangular and triangular designs, then construct the lossy transformation matrices $T$ via Eq. 6. Postselected fidelities are calculated with Eq. 7.

\begin{figure}[h]
\centering
\includegraphics[width=8.6cm, keepaspectratio]{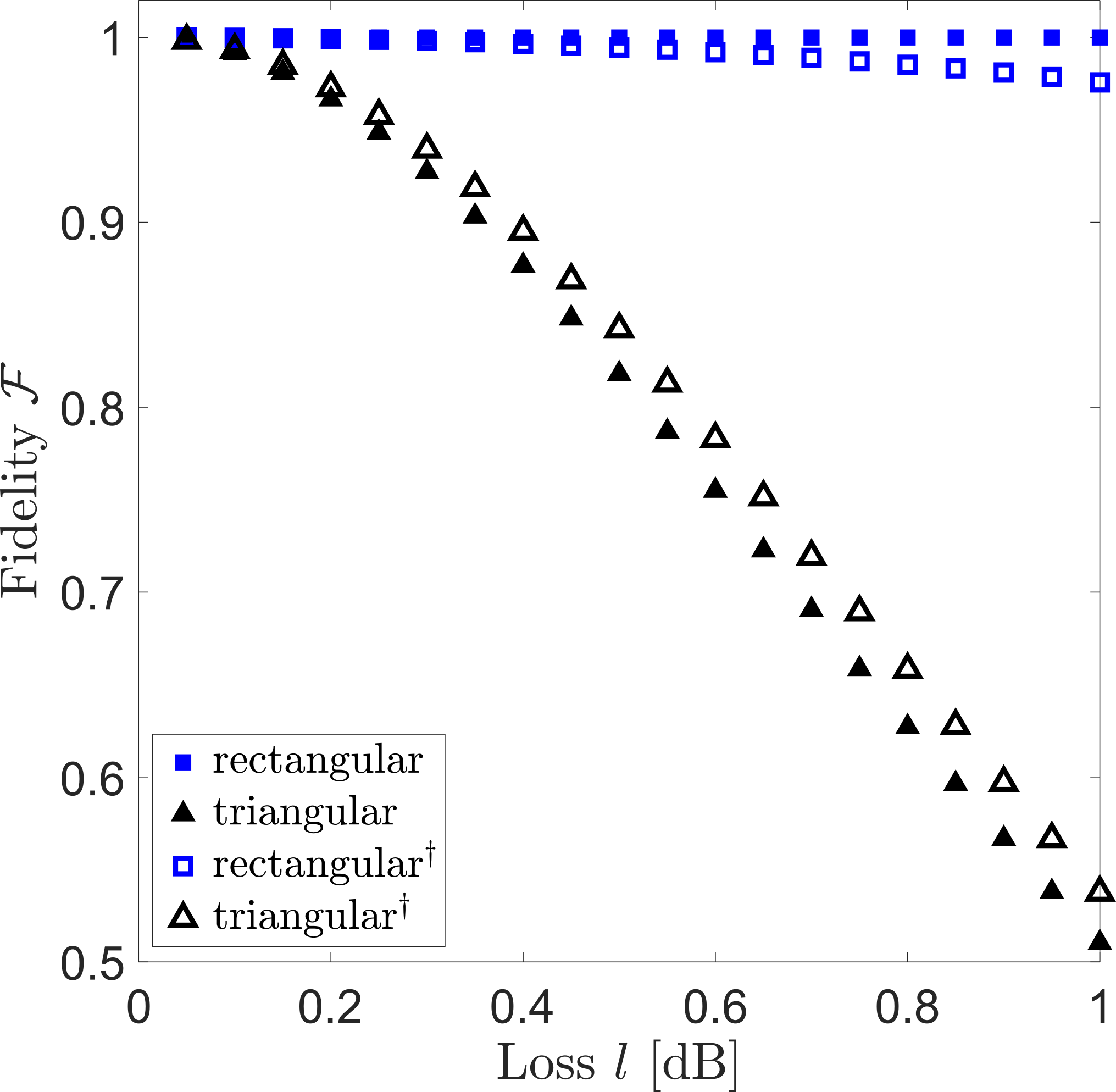}
\caption{\textbf{Postselected fidelity of 20-mode Haar-random matrices under loss.} The rectangular design retains higher fidelity under increasing losses than the triangular design. $^\dagger$Data from Ref. \cite{clements2016optimal}, where single-mode transformation losses are neglected. Points represent averages over 500 samples.}
\label{fig:FidelityInterferometer}
\end{figure}

Figure \ref{fig:FidelityInterferometer} shows that the rectangular design maintains high postselected fidelity as loss increases, while the triangular design exhibits a significant fidelity reduction. This reduction arises from asymmetric losses in the triangular design. Including losses from single-mode transformations slightly increases asymmetry in the triangular design and reduces fidelity, while the asymmetry in the rectangular design vanishes such that the fidelity is restored. 

\newpage
\section{Loss-tolerance of triangular GHZ-state generating circuits}
\label{app:GHZ3}
We analyze loss tolerance in GHZ-state generation circuits using the heralded linear optical scheme of Chen \textit{et al.} \cite{chen2024heralded} for dual-rail encoded three-qubit GHZ states. The circuit is implemented on a 12-mode rectangular multiport interferometer with a total on-chip propagation loss of 0.8 dB \cite{taballione2021universal}, corresponding to a unit cell transmission efficiency of $\eta = 10^{-0.8/120} \approx 0.9848$.

The implemented 10-mode unitary $U$ requires six photons: three for qubits and three auxiliary photons for measurement-induced nonlinearity \cite{gubarev2020improved}. Input and output permutations are arranged to minimize asymmetric loss effects in the triangular design, with six photons entering through modes 6–10 and auxiliary photons detected in output modes 1, 3, and 4. Qubits $Q_1$, $Q_2$, and $Q_3$ correspond to output mode pairs $(5,6)$, $(7,8)$, and $(9,10)$, respectively. The relevant submatrix of $U$ is given in Eq. \ref{eq:U_sub_matrix}:
\begin{equation}
\begin{split}
    U_{\text{sub}} &:= \\
    &\left [ \begin{smallmatrix}
0 & 0 & 0.3536 & -0.3536 & -0.3536 & 0.3536 \\
0 & 0 & 0.3536 & 0.3536 & 0.3536 & 0.3536 \\
0.4082 & 0.4082 & 0.2887 & 0.2887 & -0.2887 & -0.2887 \\
0.4082 & -0.4082 & 0.2887 & -0.2887 & 0.2887 & -0.2887 \\
-0.8165 & 0 & 0.2887 & 0 & 0 & -0.2887 \\
0 & 0 & 0.5000 & 0 & 0 & -0.5000 \\
0 & 0 & 0.5000 & 0 & 0 & 0.5000 \\
0 & 0.8165 & 0 & -0.2887 & 0.2887 & 0 \\
0 & 0 & 0 & -0.5000 & 0.5000 & 0 \\
0 & 0 & 0 & 0.5000 & 0.5000 & 0 \\
\end{smallmatrix} \right ],
\end{split}
\label{eq:U_sub_matrix}
\end{equation}
where columns correspond to input modes and rows correspond to output modes. Notably, a nonuniversal interferometer suffices, as photons entering specific inputs do not contribute to all qubits, leaving room for further circuit optimization beyond the triangular design.

\begin{figure}[h!]
    \centering
    \includegraphics[width=8.6cm, height=20cm, keepaspectratio,]{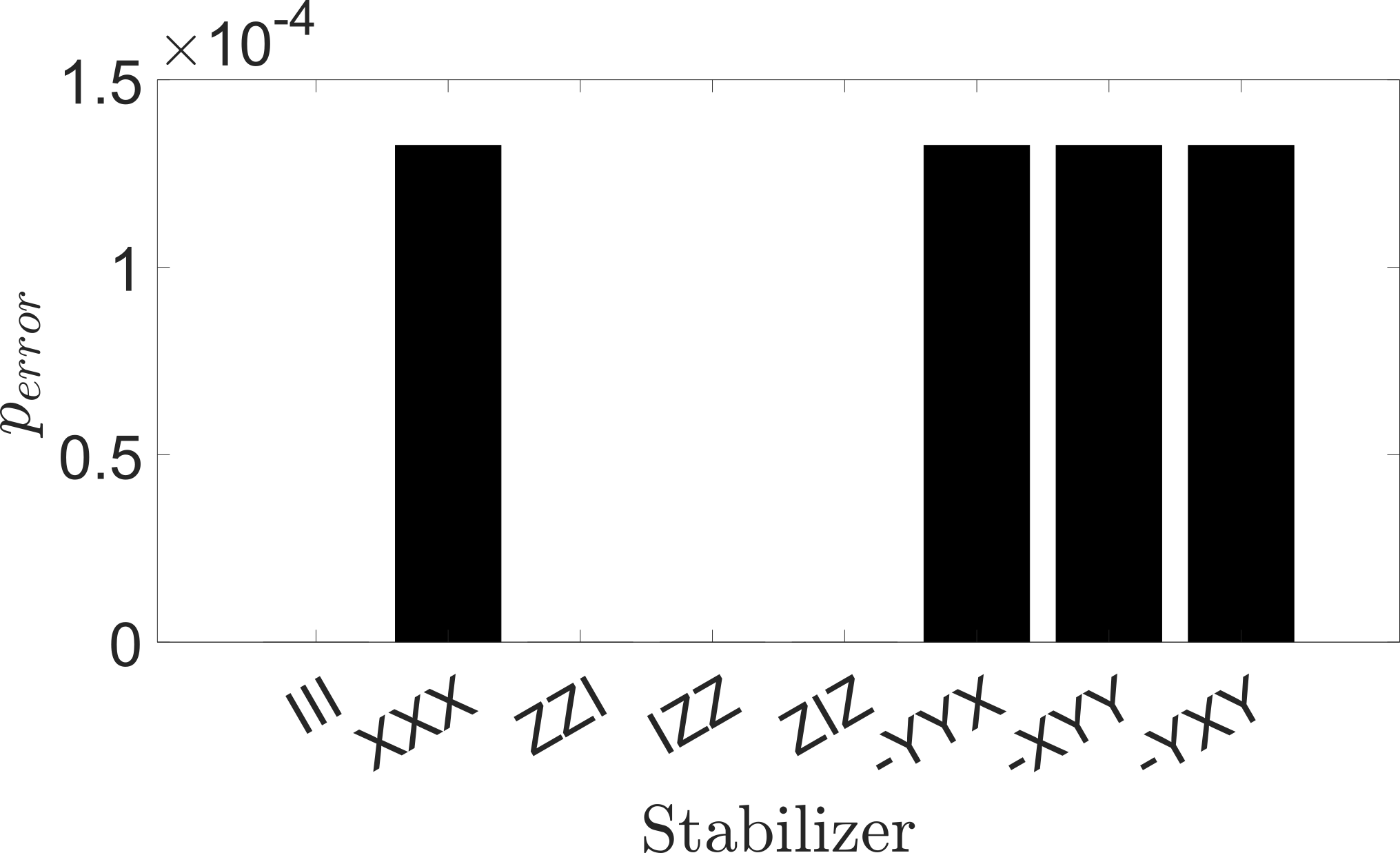}
    \caption{\textbf{Stabilizer measurement errors for the postselected heralded three-qubit GHZ state} using a triangular 10-mode interferometer with $\eta = 0.9848$. Half of the stabilizers remain error-free, half of the stabilizers are affected by an error $p_{\text{error}} = 0.013\%$.}
    \label{fig:GHZ3_stabilizererror}
\end{figure}

Switching to a triangular design can introduce stabilizer measurement errors, as the target interference is partially altered \cite{clements2016optimal}. We quantify this via
\begin{equation}
p_{\text{error}} = \frac{1}{2}\left(1 - \langle S \rangle \right), \quad \langle S \rangle = \tr{\rho^\prime S},
\end{equation}
where $\rho^\prime$ is the postselected three-qubit state and $S$ is the stabilizer operator. Fig. \ref{fig:GHZ3_stabilizererror} shows that half of the stabilizers are unaffected, while the other half exhibit $p_{\text{error}} = 0.013\%$.

The postselected state remains effectively pure under asymmetric losses \cite{clements2016optimal} and can be parametrized as
\begin{equation}
\ket{\psi^\prime} = \sin{\alpha} \ket{000} + \cos{\alpha} \ket{111}, \quad f = |\braket{\text{GHZ}_3|\psi^\prime}|^2 = \frac{1}{2} + \frac{1}{2}\sin 2\alpha.
\end{equation}
For $\eta = 0.9848$, $\alpha = 0.7969$ and $1-f = 1.3 \times 10^{-4}$.

\begin{figure}[h!]
    \centering
     \includegraphics[width=8.6cm, height=20cm, keepaspectratio,]{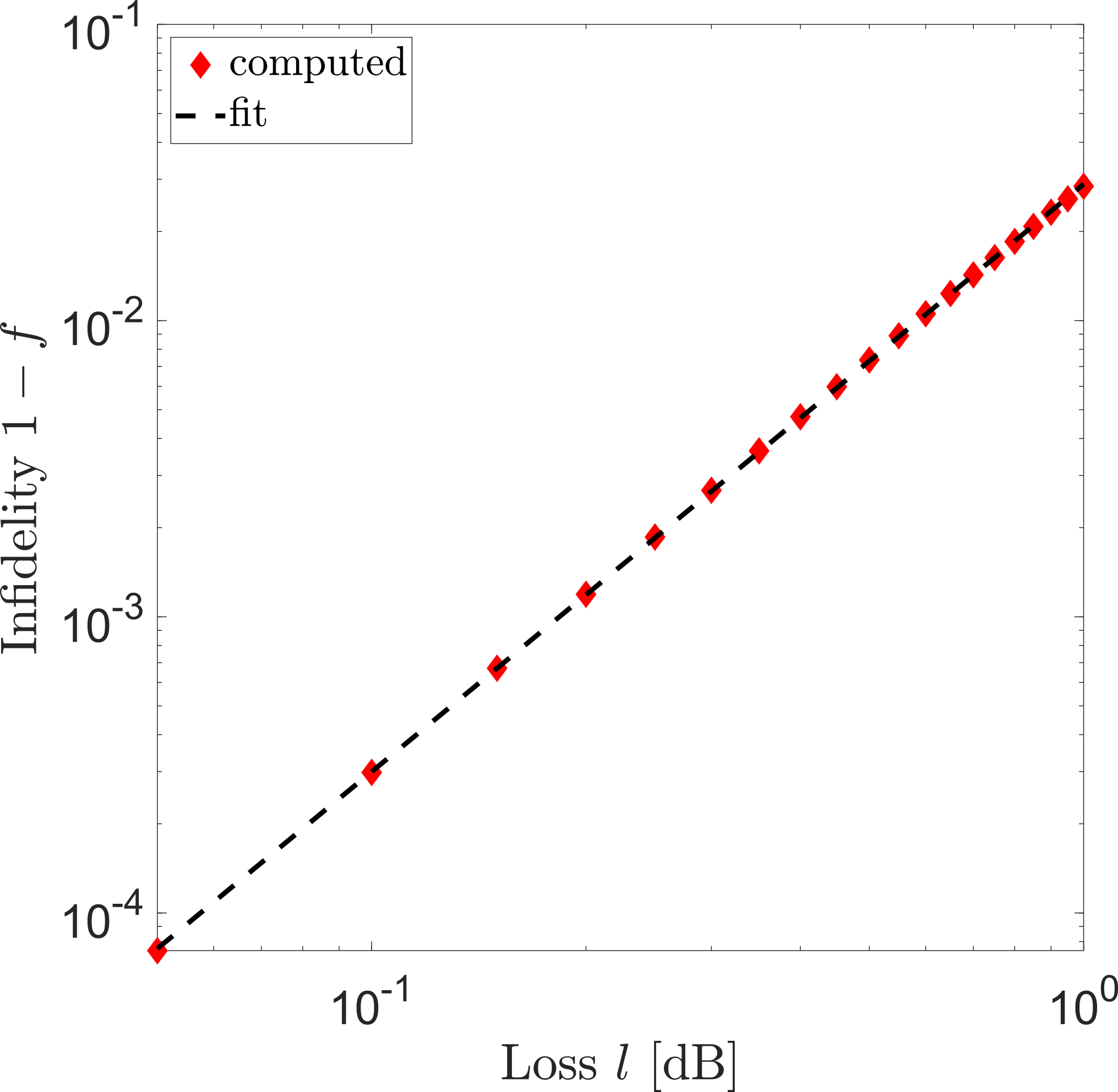}
    \caption{\textbf{Postselected GHZ-3 state infidelity} as a function of unit cell loss $l$ [dB] for a triangular interferometer. The infidelity scales approximately quadratically with loss, fitted as $1-f = 0.0289 \cdot l^{1.984}$.}
    \label{fig:FidelityGHZ3}
\end{figure}

Fig. \ref{fig:FidelityGHZ3} shows that the postselected state infidelity scales approximately quadratically with unit cell loss, indicating that stabilizer errors due to asymmetric losses in a triangular design are a second-order effect.

\newpage
\bibliography{refs.bib}

\begin{thebibliography}{58}%
\makeatletter
\providecommand \@ifxundefined [1]{%
 \@ifx{#1\undefined}
}%
\providecommand \@ifnum [1]{%
 \ifnum #1\expandafter \@firstoftwo
 \else \expandafter \@secondoftwo
 \fi
}%
\providecommand \@ifx [1]{%
 \ifx #1\expandafter \@firstoftwo
 \else \expandafter \@secondoftwo
 \fi
}%
\providecommand \natexlab [1]{#1}%
\providecommand \enquote  [1]{``#1''}%
\providecommand \bibnamefont  [1]{#1}%
\providecommand \bibfnamefont [1]{#1}%
\providecommand \citenamefont [1]{#1}%
\providecommand \href@noop [0]{\@secondoftwo}%
\providecommand \href [0]{\begingroup \@sanitize@url \@href}%
\providecommand \@href[1]{\@@startlink{#1}\@@href}%
\providecommand \@@href[1]{\endgroup#1\@@endlink}%
\providecommand \@sanitize@url [0]{\catcode `\\12\catcode `\$12\catcode `\&12\catcode `\#12\catcode `\^12\catcode `\_12\catcode `\%12\relax}%
\providecommand \@@startlink[1]{}%
\providecommand \@@endlink[0]{}%
\providecommand \url  [0]{\begingroup\@sanitize@url \@url }%
\providecommand \@url [1]{\endgroup\@href {#1}{\urlprefix }}%
\providecommand \urlprefix  [0]{URL }%
\providecommand \Eprint [0]{\href }%
\providecommand \doibase [0]{https://doi.org/}%
\providecommand \selectlanguage [0]{\@gobble}%
\providecommand \bibinfo  [0]{\@secondoftwo}%
\providecommand \bibfield  [0]{\@secondoftwo}%
\providecommand \translation [1]{[#1]}%
\providecommand \BibitemOpen [0]{}%
\providecommand \bibitemStop [0]{}%
\providecommand \bibitemNoStop [0]{.\EOS\space}%
\providecommand \EOS [0]{\spacefactor3000\relax}%
\providecommand \BibitemShut  [1]{\csname bibitem#1\endcsname}%
\let\auto@bib@innerbib\@empty
\bibitem [{\citenamefont {Carolan}\ \emph {et~al.}(2015)\citenamefont {Carolan}, \citenamefont {Harrold}, \citenamefont {Sparrow}, \citenamefont {Mart{\'\i}n-L{\'o}pez}, \citenamefont {Russell}, \citenamefont {Silverstone}, \citenamefont {Shadbolt}, \citenamefont {Matsuda}, \citenamefont {Oguma}, \citenamefont {Itoh} \emph {et~al.}}]{carolan2015universal}%
  \BibitemOpen
  \bibfield  {author} {\bibinfo {author} {\bibfnamefont {J.}~\bibnamefont {Carolan}}, \bibinfo {author} {\bibfnamefont {C.}~\bibnamefont {Harrold}}, \bibinfo {author} {\bibfnamefont {C.}~\bibnamefont {Sparrow}}, \bibinfo {author} {\bibfnamefont {E.}~\bibnamefont {Mart{\'\i}n-L{\'o}pez}}, \bibinfo {author} {\bibfnamefont {N.~J.}\ \bibnamefont {Russell}}, \bibinfo {author} {\bibfnamefont {J.~W.}\ \bibnamefont {Silverstone}}, \bibinfo {author} {\bibfnamefont {P.~J.}\ \bibnamefont {Shadbolt}}, \bibinfo {author} {\bibfnamefont {N.}~\bibnamefont {Matsuda}}, \bibinfo {author} {\bibfnamefont {M.}~\bibnamefont {Oguma}}, \bibinfo {author} {\bibfnamefont {M.}~\bibnamefont {Itoh}}, \emph {et~al.},\ }\bibfield  {title} {\bibinfo {title} {Universal linear optics},\ }\href@noop {} {\bibfield  {journal} {\bibinfo  {journal} {Science}\ }\textbf {\bibinfo {volume} {349}},\ \bibinfo {pages} {711} (\bibinfo {year} {2015})}\BibitemShut {NoStop}%
\bibitem [{\citenamefont {Adcock}\ \emph {et~al.}(2019)\citenamefont {Adcock}, \citenamefont {Vigliar}, \citenamefont {Santagati}, \citenamefont {Silverstone},\ and\ \citenamefont {Thompson}}]{adcock2019programmable}%
  \BibitemOpen
  \bibfield  {author} {\bibinfo {author} {\bibfnamefont {J.~C.}\ \bibnamefont {Adcock}}, \bibinfo {author} {\bibfnamefont {C.}~\bibnamefont {Vigliar}}, \bibinfo {author} {\bibfnamefont {R.}~\bibnamefont {Santagati}}, \bibinfo {author} {\bibfnamefont {J.~W.}\ \bibnamefont {Silverstone}},\ and\ \bibinfo {author} {\bibfnamefont {M.~G.}\ \bibnamefont {Thompson}},\ }\bibfield  {title} {\bibinfo {title} {Programmable four-photon graph states on a silicon chip},\ }\href@noop {} {\bibfield  {journal} {\bibinfo  {journal} {Nature communications}\ }\textbf {\bibinfo {volume} {10}},\ \bibinfo {pages} {3528} (\bibinfo {year} {2019})}\BibitemShut {NoStop}%
\bibitem [{\citenamefont {Llewellyn}\ \emph {et~al.}(2020)\citenamefont {Llewellyn}, \citenamefont {Ding}, \citenamefont {Faruque}, \citenamefont {Paesani}, \citenamefont {Bacco}, \citenamefont {Santagati}, \citenamefont {Qian}, \citenamefont {Li}, \citenamefont {Xiao}, \citenamefont {Huber} \emph {et~al.}}]{llewellyn2020chip}%
  \BibitemOpen
  \bibfield  {author} {\bibinfo {author} {\bibfnamefont {D.}~\bibnamefont {Llewellyn}}, \bibinfo {author} {\bibfnamefont {Y.}~\bibnamefont {Ding}}, \bibinfo {author} {\bibfnamefont {I.~I.}\ \bibnamefont {Faruque}}, \bibinfo {author} {\bibfnamefont {S.}~\bibnamefont {Paesani}}, \bibinfo {author} {\bibfnamefont {D.}~\bibnamefont {Bacco}}, \bibinfo {author} {\bibfnamefont {R.}~\bibnamefont {Santagati}}, \bibinfo {author} {\bibfnamefont {Y.-J.}\ \bibnamefont {Qian}}, \bibinfo {author} {\bibfnamefont {Y.}~\bibnamefont {Li}}, \bibinfo {author} {\bibfnamefont {Y.-F.}\ \bibnamefont {Xiao}}, \bibinfo {author} {\bibfnamefont {M.}~\bibnamefont {Huber}}, \emph {et~al.},\ }\bibfield  {title} {\bibinfo {title} {Chip-to-chip quantum teleportation and multi-photon entanglement in silicon},\ }\href@noop {} {\bibfield  {journal} {\bibinfo  {journal} {Nature Physics}\ }\textbf {\bibinfo {volume} {16}},\ \bibinfo {pages} {148} (\bibinfo {year} {2020})}\BibitemShut {NoStop}%
\bibitem [{\citenamefont {Li}\ \emph {et~al.}(2022)\citenamefont {Li}, \citenamefont {Wan}, \citenamefont {Zhang}, \citenamefont {Zhu}, \citenamefont {Shi}, \citenamefont {Chin}, \citenamefont {Zhou}, \citenamefont {Kwek},\ and\ \citenamefont {Liu}}]{li2022quantum}%
  \BibitemOpen
  \bibfield  {author} {\bibinfo {author} {\bibfnamefont {Y.}~\bibnamefont {Li}}, \bibinfo {author} {\bibfnamefont {L.}~\bibnamefont {Wan}}, \bibinfo {author} {\bibfnamefont {H.}~\bibnamefont {Zhang}}, \bibinfo {author} {\bibfnamefont {H.}~\bibnamefont {Zhu}}, \bibinfo {author} {\bibfnamefont {Y.}~\bibnamefont {Shi}}, \bibinfo {author} {\bibfnamefont {L.~K.}\ \bibnamefont {Chin}}, \bibinfo {author} {\bibfnamefont {X.}~\bibnamefont {Zhou}}, \bibinfo {author} {\bibfnamefont {L.~C.}\ \bibnamefont {Kwek}},\ and\ \bibinfo {author} {\bibfnamefont {A.~Q.}\ \bibnamefont {Liu}},\ }\bibfield  {title} {\bibinfo {title} {Quantum fredkin and toffoli gates on a versatile programmable silicon photonic chip},\ }\href@noop {} {\bibfield  {journal} {\bibinfo  {journal} {npj Quantum Information}\ }\textbf {\bibinfo {volume} {8}},\ \bibinfo {pages} {112} (\bibinfo {year} {2022})}\BibitemShut {NoStop}%
\bibitem [{\citenamefont {Moody}\ \emph {et~al.}(2022)\citenamefont {Moody}, \citenamefont {Sorger}, \citenamefont {Blumenthal}, \citenamefont {Juodawlkis}, \citenamefont {Loh}, \citenamefont {Sorace-Agaskar}, \citenamefont {Jones}, \citenamefont {Balram}, \citenamefont {Matthews}, \citenamefont {Laing} \emph {et~al.}}]{moody2022}%
  \BibitemOpen
  \bibfield  {author} {\bibinfo {author} {\bibfnamefont {G.}~\bibnamefont {Moody}}, \bibinfo {author} {\bibfnamefont {V.~J.}\ \bibnamefont {Sorger}}, \bibinfo {author} {\bibfnamefont {D.~J.}\ \bibnamefont {Blumenthal}}, \bibinfo {author} {\bibfnamefont {P.~W.}\ \bibnamefont {Juodawlkis}}, \bibinfo {author} {\bibfnamefont {W.}~\bibnamefont {Loh}}, \bibinfo {author} {\bibfnamefont {C.}~\bibnamefont {Sorace-Agaskar}}, \bibinfo {author} {\bibfnamefont {A.~E.}\ \bibnamefont {Jones}}, \bibinfo {author} {\bibfnamefont {K.~C.}\ \bibnamefont {Balram}}, \bibinfo {author} {\bibfnamefont {J.~C.}\ \bibnamefont {Matthews}}, \bibinfo {author} {\bibfnamefont {A.}~\bibnamefont {Laing}}, \emph {et~al.},\ }\bibfield  {title} {\bibinfo {title} {2022 roadmap on integrated quantum photonics},\ }\href@noop {} {\bibfield  {journal} {\bibinfo  {journal} {Journal of Physics: Photonics}\ }\textbf {\bibinfo {volume} {4}},\ \bibinfo {pages} {012501} (\bibinfo {year} {2022})}\BibitemShut {NoStop}%
\bibitem [{\citenamefont {Pai}\ \emph {et~al.}(2023)\citenamefont {Pai}, \citenamefont {Park}, \citenamefont {Ball}, \citenamefont {Penkovsky}, \citenamefont {Dubrovsky}, \citenamefont {Abebe}, \citenamefont {Milanizadeh}, \citenamefont {Morichetti}, \citenamefont {Melloni}, \citenamefont {Fan} \emph {et~al.}}]{pai2023experimental}%
  \BibitemOpen
  \bibfield  {author} {\bibinfo {author} {\bibfnamefont {S.}~\bibnamefont {Pai}}, \bibinfo {author} {\bibfnamefont {T.}~\bibnamefont {Park}}, \bibinfo {author} {\bibfnamefont {M.}~\bibnamefont {Ball}}, \bibinfo {author} {\bibfnamefont {B.}~\bibnamefont {Penkovsky}}, \bibinfo {author} {\bibfnamefont {M.}~\bibnamefont {Dubrovsky}}, \bibinfo {author} {\bibfnamefont {N.}~\bibnamefont {Abebe}}, \bibinfo {author} {\bibfnamefont {M.}~\bibnamefont {Milanizadeh}}, \bibinfo {author} {\bibfnamefont {F.}~\bibnamefont {Morichetti}}, \bibinfo {author} {\bibfnamefont {A.}~\bibnamefont {Melloni}}, \bibinfo {author} {\bibfnamefont {S.}~\bibnamefont {Fan}}, \emph {et~al.},\ }\bibfield  {title} {\bibinfo {title} {Experimental evaluation of digitally verifiable photonic computing for blockchain and cryptocurrency},\ }\href@noop {} {\bibfield  {journal} {\bibinfo  {journal} {Optica}\ }\textbf {\bibinfo {volume} {10}},\ \bibinfo {pages} {552} (\bibinfo {year} {2023})}\BibitemShut {NoStop}%
\bibitem [{\citenamefont {Huang}\ \emph {et~al.}(2024)\citenamefont {Huang}, \citenamefont {Li}, \citenamefont {Chen}, \citenamefont {Zhai}, \citenamefont {Zheng}, \citenamefont {Chi}, \citenamefont {Li}, \citenamefont {He}, \citenamefont {Gong},\ and\ \citenamefont {Wang}}]{huang2024demonstration}%
  \BibitemOpen
  \bibfield  {author} {\bibinfo {author} {\bibfnamefont {J.}~\bibnamefont {Huang}}, \bibinfo {author} {\bibfnamefont {X.}~\bibnamefont {Li}}, \bibinfo {author} {\bibfnamefont {X.}~\bibnamefont {Chen}}, \bibinfo {author} {\bibfnamefont {C.}~\bibnamefont {Zhai}}, \bibinfo {author} {\bibfnamefont {Y.}~\bibnamefont {Zheng}}, \bibinfo {author} {\bibfnamefont {Y.}~\bibnamefont {Chi}}, \bibinfo {author} {\bibfnamefont {Y.}~\bibnamefont {Li}}, \bibinfo {author} {\bibfnamefont {Q.}~\bibnamefont {He}}, \bibinfo {author} {\bibfnamefont {Q.}~\bibnamefont {Gong}},\ and\ \bibinfo {author} {\bibfnamefont {J.}~\bibnamefont {Wang}},\ }\bibfield  {title} {\bibinfo {title} {Demonstration of hypergraph-state quantum information processing},\ }\href@noop {} {\bibfield  {journal} {\bibinfo  {journal} {Nature Communications}\ }\textbf {\bibinfo {volume} {15}},\ \bibinfo {pages} {2601} (\bibinfo {year} {2024})}\BibitemShut {NoStop}%
\bibitem [{\citenamefont {Hoch}\ \emph {et~al.}(2025)\citenamefont {Hoch}, \citenamefont {Caruccio}, \citenamefont {Rodari}, \citenamefont {Francalanci}, \citenamefont {Suprano}, \citenamefont {Giordani}, \citenamefont {Carvacho}, \citenamefont {Spagnolo}, \citenamefont {Koudia}, \citenamefont {Proietti} \emph {et~al.}}]{hoch2025quantum}%
  \BibitemOpen
  \bibfield  {author} {\bibinfo {author} {\bibfnamefont {F.}~\bibnamefont {Hoch}}, \bibinfo {author} {\bibfnamefont {E.}~\bibnamefont {Caruccio}}, \bibinfo {author} {\bibfnamefont {G.}~\bibnamefont {Rodari}}, \bibinfo {author} {\bibfnamefont {T.}~\bibnamefont {Francalanci}}, \bibinfo {author} {\bibfnamefont {A.}~\bibnamefont {Suprano}}, \bibinfo {author} {\bibfnamefont {T.}~\bibnamefont {Giordani}}, \bibinfo {author} {\bibfnamefont {G.}~\bibnamefont {Carvacho}}, \bibinfo {author} {\bibfnamefont {N.}~\bibnamefont {Spagnolo}}, \bibinfo {author} {\bibfnamefont {S.}~\bibnamefont {Koudia}}, \bibinfo {author} {\bibfnamefont {M.}~\bibnamefont {Proietti}}, \emph {et~al.},\ }\bibfield  {title} {\bibinfo {title} {Quantum machine learning with adaptive boson sampling via post-selection},\ }\href@noop {} {\bibfield  {journal} {\bibinfo  {journal} {Nature Communications}\ }\textbf {\bibinfo {volume} {16}},\ \bibinfo {pages} {902} (\bibinfo {year} {2025})}\BibitemShut {NoStop}%
\bibitem [{\citenamefont {Li}\ \emph {et~al.}(2025)\citenamefont {Li}, \citenamefont {Zhu}, \citenamefont {Luo}, \citenamefont {Cai}, \citenamefont {Karim}, \citenamefont {Luo}, \citenamefont {Gao}, \citenamefont {Wu}, \citenamefont {Zhou}, \citenamefont {Song} \emph {et~al.}}]{li2025realizing}%
  \BibitemOpen
  \bibfield  {author} {\bibinfo {author} {\bibfnamefont {Y.}~\bibnamefont {Li}}, \bibinfo {author} {\bibfnamefont {H.}~\bibnamefont {Zhu}}, \bibinfo {author} {\bibfnamefont {W.}~\bibnamefont {Luo}}, \bibinfo {author} {\bibfnamefont {H.}~\bibnamefont {Cai}}, \bibinfo {author} {\bibfnamefont {M.~F.}\ \bibnamefont {Karim}}, \bibinfo {author} {\bibfnamefont {X.}~\bibnamefont {Luo}}, \bibinfo {author} {\bibfnamefont {F.}~\bibnamefont {Gao}}, \bibinfo {author} {\bibfnamefont {X.}~\bibnamefont {Wu}}, \bibinfo {author} {\bibfnamefont {X.}~\bibnamefont {Zhou}}, \bibinfo {author} {\bibfnamefont {Q.}~\bibnamefont {Song}}, \emph {et~al.},\ }\bibfield  {title} {\bibinfo {title} {Realizing ultrahigh capacity quantum superdense coding on quantum photonic chip},\ }\href@noop {} {\bibfield  {journal} {\bibinfo  {journal} {npj Quantum Information}\ }\textbf {\bibinfo {volume} {11}},\ \bibinfo {pages} {49} (\bibinfo {year} {2025})}\BibitemShut {NoStop}%
\bibitem [{\citenamefont {Salavrakos}\ \emph {et~al.}(2025)\citenamefont {Salavrakos}, \citenamefont {Maring}, \citenamefont {Emeriau},\ and\ \citenamefont {Mansfield}}]{salavrakos2025photon}%
  \BibitemOpen
  \bibfield  {author} {\bibinfo {author} {\bibfnamefont {A.}~\bibnamefont {Salavrakos}}, \bibinfo {author} {\bibfnamefont {N.}~\bibnamefont {Maring}}, \bibinfo {author} {\bibfnamefont {P.-E.}\ \bibnamefont {Emeriau}},\ and\ \bibinfo {author} {\bibfnamefont {S.}~\bibnamefont {Mansfield}},\ }\bibfield  {title} {\bibinfo {title} {Photon-native quantum algorithms},\ }\href@noop {} {\bibfield  {journal} {\bibinfo  {journal} {Materials for Quantum Technology}\ } (\bibinfo {year} {2025})}\BibitemShut {NoStop}%
\bibitem [{psi(2025)}]{psiquantum2025manufacturable}%
  \BibitemOpen
  \bibfield  {title} {\bibinfo {title} {A manufacturable platform for photonic quantum computing},\ }\href@noop {} {\bibfield  {journal} {\bibinfo  {journal} {Nature}\ ,\ \bibinfo {pages} {1}} (\bibinfo {year} {2025})}\BibitemShut {NoStop}%
\bibitem [{\citenamefont {Miller}(1969)}]{miller1969integrated}%
  \BibitemOpen
  \bibfield  {author} {\bibinfo {author} {\bibfnamefont {S.~E.}\ \bibnamefont {Miller}},\ }\bibfield  {title} {\bibinfo {title} {Integrated optics: An introduction},\ }\href@noop {} {\bibfield  {journal} {\bibinfo  {journal} {The Bell system technical journal}\ }\textbf {\bibinfo {volume} {48}},\ \bibinfo {pages} {2059} (\bibinfo {year} {1969})}\BibitemShut {NoStop}%
\bibitem [{\citenamefont {Nielsen}\ and\ \citenamefont {Chuang}(2010)}]{nielsen2010quantum}%
  \BibitemOpen
  \bibfield  {author} {\bibinfo {author} {\bibfnamefont {M.~A.}\ \bibnamefont {Nielsen}}\ and\ \bibinfo {author} {\bibfnamefont {I.~L.}\ \bibnamefont {Chuang}},\ }\href@noop {} {\emph {\bibinfo {title} {Quantum computation and quantum information}}}\ (\bibinfo  {publisher} {Cambridge university press},\ \bibinfo {year} {2010})\BibitemShut {NoStop}%
\bibitem [{\citenamefont {Reck}\ \emph {et~al.}(1994)\citenamefont {Reck}, \citenamefont {Zeilinger}, \citenamefont {Bernstein},\ and\ \citenamefont {Bertani}}]{reck1994experimental}%
  \BibitemOpen
  \bibfield  {author} {\bibinfo {author} {\bibfnamefont {M.}~\bibnamefont {Reck}}, \bibinfo {author} {\bibfnamefont {A.}~\bibnamefont {Zeilinger}}, \bibinfo {author} {\bibfnamefont {H.~J.}\ \bibnamefont {Bernstein}},\ and\ \bibinfo {author} {\bibfnamefont {P.}~\bibnamefont {Bertani}},\ }\bibfield  {title} {\bibinfo {title} {Experimental realization of any discrete unitary operator},\ }\href@noop {} {\bibfield  {journal} {\bibinfo  {journal} {Physical review letters}\ }\textbf {\bibinfo {volume} {73}},\ \bibinfo {pages} {58} (\bibinfo {year} {1994})}\BibitemShut {NoStop}%
\bibitem [{\citenamefont {Clements}\ \emph {et~al.}(2016)\citenamefont {Clements}, \citenamefont {Humphreys}, \citenamefont {Metcalf}, \citenamefont {Kolthammer},\ and\ \citenamefont {Walmsley}}]{clements2016optimal}%
  \BibitemOpen
  \bibfield  {author} {\bibinfo {author} {\bibfnamefont {W.~R.}\ \bibnamefont {Clements}}, \bibinfo {author} {\bibfnamefont {P.~C.}\ \bibnamefont {Humphreys}}, \bibinfo {author} {\bibfnamefont {B.~J.}\ \bibnamefont {Metcalf}}, \bibinfo {author} {\bibfnamefont {W.~S.}\ \bibnamefont {Kolthammer}},\ and\ \bibinfo {author} {\bibfnamefont {I.~A.}\ \bibnamefont {Walmsley}},\ }\bibfield  {title} {\bibinfo {title} {Optimal design for universal multiport interferometers},\ }\href@noop {} {\bibfield  {journal} {\bibinfo  {journal} {Optica}\ }\textbf {\bibinfo {volume} {3}},\ \bibinfo {pages} {1460} (\bibinfo {year} {2016})}\BibitemShut {NoStop}%
\bibitem [{\citenamefont {Vandebril}(2011)}]{vandebril2011chasing}%
  \BibitemOpen
  \bibfield  {author} {\bibinfo {author} {\bibfnamefont {R.}~\bibnamefont {Vandebril}},\ }\bibfield  {title} {\bibinfo {title} {Chasing bulges or rotations? a metamorphosis of the qr-algorithm},\ }\href@noop {} {\bibfield  {journal} {\bibinfo  {journal} {SIAM Journal on Matrix Analysis and Applications}\ }\textbf {\bibinfo {volume} {32}},\ \bibinfo {pages} {217} (\bibinfo {year} {2011})}\BibitemShut {NoStop}%
\bibitem [{\citenamefont {Cl{\'e}ment}\ \emph {et~al.}(2022)\citenamefont {Cl{\'e}ment}, \citenamefont {Heurtel}, \citenamefont {Mansfield}, \citenamefont {Perdrix},\ and\ \citenamefont {Valiron}}]{clement2022lov}%
  \BibitemOpen
  \bibfield  {author} {\bibinfo {author} {\bibfnamefont {A.}~\bibnamefont {Cl{\'e}ment}}, \bibinfo {author} {\bibfnamefont {N.}~\bibnamefont {Heurtel}}, \bibinfo {author} {\bibfnamefont {S.}~\bibnamefont {Mansfield}}, \bibinfo {author} {\bibfnamefont {S.}~\bibnamefont {Perdrix}},\ and\ \bibinfo {author} {\bibfnamefont {B.}~\bibnamefont {Valiron}},\ }\bibfield  {title} {\bibinfo {title} {Lov-calculus: A graphical language for linear optical quantum circuits},\ }\href@noop {} {\bibfield  {journal} {\bibinfo  {journal} {arXiv preprint arXiv:2204.11787}\ } (\bibinfo {year} {2022})}\BibitemShut {NoStop}%
\bibitem [{\citenamefont {Knill}\ \emph {et~al.}(2001)\citenamefont {Knill}, \citenamefont {Laflamme},\ and\ \citenamefont {Milburn}}]{knill2001scheme}%
  \BibitemOpen
  \bibfield  {author} {\bibinfo {author} {\bibfnamefont {E.}~\bibnamefont {Knill}}, \bibinfo {author} {\bibfnamefont {R.}~\bibnamefont {Laflamme}},\ and\ \bibinfo {author} {\bibfnamefont {G.~J.}\ \bibnamefont {Milburn}},\ }\bibfield  {title} {\bibinfo {title} {A scheme for efficient quantum computation with linear optics},\ }\href@noop {} {\bibfield  {journal} {\bibinfo  {journal} {nature}\ }\textbf {\bibinfo {volume} {409}},\ \bibinfo {pages} {46} (\bibinfo {year} {2001})}\BibitemShut {NoStop}%
\bibitem [{\citenamefont {Aaronson}\ and\ \citenamefont {Arkhipov}(2011)}]{aaronson2011computational}%
  \BibitemOpen
  \bibfield  {author} {\bibinfo {author} {\bibfnamefont {S.}~\bibnamefont {Aaronson}}\ and\ \bibinfo {author} {\bibfnamefont {A.}~\bibnamefont {Arkhipov}},\ }\bibfield  {title} {\bibinfo {title} {The computational complexity of linear optics},\ }in\ \href@noop {} {\emph {\bibinfo {booktitle} {Proceedings of the forty-third annual ACM symposium on Theory of computing}}}\ (\bibinfo {year} {2011})\ pp.\ \bibinfo {pages} {333--342}\BibitemShut {NoStop}%
\bibitem [{\citenamefont {Brod}\ \emph {et~al.}(2019)\citenamefont {Brod}, \citenamefont {Galv{\~a}o}, \citenamefont {Crespi}, \citenamefont {Osellame}, \citenamefont {Spagnolo},\ and\ \citenamefont {Sciarrino}}]{brod2019photonic}%
  \BibitemOpen
  \bibfield  {author} {\bibinfo {author} {\bibfnamefont {D.~J.}\ \bibnamefont {Brod}}, \bibinfo {author} {\bibfnamefont {E.~F.}\ \bibnamefont {Galv{\~a}o}}, \bibinfo {author} {\bibfnamefont {A.}~\bibnamefont {Crespi}}, \bibinfo {author} {\bibfnamefont {R.}~\bibnamefont {Osellame}}, \bibinfo {author} {\bibfnamefont {N.}~\bibnamefont {Spagnolo}},\ and\ \bibinfo {author} {\bibfnamefont {F.}~\bibnamefont {Sciarrino}},\ }\bibfield  {title} {\bibinfo {title} {Photonic implementation of boson sampling: a review},\ }\href@noop {} {\bibfield  {journal} {\bibinfo  {journal} {Advanced Photonics}\ }\textbf {\bibinfo {volume} {1}},\ \bibinfo {pages} {034001} (\bibinfo {year} {2019})}\BibitemShut {NoStop}%
\bibitem [{\citenamefont {Forbes}\ \emph {et~al.}(2025)\citenamefont {Forbes}, \citenamefont {Ghafari}, \citenamefont {Deacon}, \citenamefont {Singh}, \citenamefont {Lavie}, \citenamefont {Yard}, \citenamefont {Shaw}, \citenamefont {Laing},\ and\ \citenamefont {Tischler}}]{forbes2025heralded}%
  \BibitemOpen
  \bibfield  {author} {\bibinfo {author} {\bibfnamefont {I.}~\bibnamefont {Forbes}}, \bibinfo {author} {\bibfnamefont {F.}~\bibnamefont {Ghafari}}, \bibinfo {author} {\bibfnamefont {E.}~\bibnamefont {Deacon}}, \bibinfo {author} {\bibfnamefont {S.~P.~P.}\ \bibnamefont {Singh}}, \bibinfo {author} {\bibfnamefont {E.}~\bibnamefont {Lavie}}, \bibinfo {author} {\bibfnamefont {P.}~\bibnamefont {Yard}}, \bibinfo {author} {\bibfnamefont {R.}~\bibnamefont {Shaw}}, \bibinfo {author} {\bibfnamefont {A.}~\bibnamefont {Laing}},\ and\ \bibinfo {author} {\bibfnamefont {N.}~\bibnamefont {Tischler}},\ }\bibfield  {title} {\bibinfo {title} {Heralded generation of entanglement with photons},\ }\href@noop {} {\bibfield  {journal} {\bibinfo  {journal} {Reports on Progress in Physics}\ } (\bibinfo {year} {2025})}\BibitemShut {NoStop}%
\bibitem [{\citenamefont {Gubarev}\ \emph {et~al.}(2020)\citenamefont {Gubarev}, \citenamefont {Dyakonov}, \citenamefont {Saygin}, \citenamefont {Struchalin}, \citenamefont {Straupe},\ and\ \citenamefont {Kulik}}]{gubarev2020improved}%
  \BibitemOpen
  \bibfield  {author} {\bibinfo {author} {\bibfnamefont {F.}~\bibnamefont {Gubarev}}, \bibinfo {author} {\bibfnamefont {I.}~\bibnamefont {Dyakonov}}, \bibinfo {author} {\bibfnamefont {M.~Y.}\ \bibnamefont {Saygin}}, \bibinfo {author} {\bibfnamefont {G.}~\bibnamefont {Struchalin}}, \bibinfo {author} {\bibfnamefont {S.}~\bibnamefont {Straupe}},\ and\ \bibinfo {author} {\bibfnamefont {S.}~\bibnamefont {Kulik}},\ }\bibfield  {title} {\bibinfo {title} {Improved heralded schemes to generate entangled states from single photons},\ }\href@noop {} {\bibfield  {journal} {\bibinfo  {journal} {Physical Review A}\ }\textbf {\bibinfo {volume} {102}},\ \bibinfo {pages} {012604} (\bibinfo {year} {2020})}\BibitemShut {NoStop}%
\bibitem [{\citenamefont {Marshall}(2022)}]{marshall2022distillation}%
  \BibitemOpen
  \bibfield  {author} {\bibinfo {author} {\bibfnamefont {J.}~\bibnamefont {Marshall}},\ }\bibfield  {title} {\bibinfo {title} {Distillation of indistinguishable photons},\ }\href@noop {} {\bibfield  {journal} {\bibinfo  {journal} {Physical Review Letters}\ }\textbf {\bibinfo {volume} {129}},\ \bibinfo {pages} {213601} (\bibinfo {year} {2022})}\BibitemShut {NoStop}%
\bibitem [{\citenamefont {Somhorst}\ \emph {et~al.}(2025)\citenamefont {Somhorst}, \citenamefont {Sauër}, \citenamefont {van~den Hoven},\ and\ \citenamefont {Renema}}]{somhorst2024photon}%
  \BibitemOpen
  \bibfield  {author} {\bibinfo {author} {\bibfnamefont {F.~H.~B.}\ \bibnamefont {Somhorst}}, \bibinfo {author} {\bibfnamefont {B.~K.}\ \bibnamefont {Sauër}}, \bibinfo {author} {\bibfnamefont {S.~N.}\ \bibnamefont {van~den Hoven}},\ and\ \bibinfo {author} {\bibfnamefont {J.~J.}\ \bibnamefont {Renema}},\ }\bibfield  {title} {\bibinfo {title} {Photon-distillation schemes with reduced resource costs based on multiphoton fourier interference},\ }\href@noop {} {\bibfield  {journal} {\bibinfo  {journal} {Physical Review Applied}\ }\textbf {\bibinfo {volume} {23}},\ \bibinfo {pages} {044003} (\bibinfo {year} {2025})}\BibitemShut {NoStop}%
\bibitem [{\citenamefont {Saied}\ \emph {et~al.}(2025)\citenamefont {Saied}, \citenamefont {Marshall}, \citenamefont {Anand},\ and\ \citenamefont {Rieffel}}]{saied2024general}%
  \BibitemOpen
  \bibfield  {author} {\bibinfo {author} {\bibfnamefont {J.}~\bibnamefont {Saied}}, \bibinfo {author} {\bibfnamefont {J.}~\bibnamefont {Marshall}}, \bibinfo {author} {\bibfnamefont {N.}~\bibnamefont {Anand}},\ and\ \bibinfo {author} {\bibfnamefont {E.~G.}\ \bibnamefont {Rieffel}},\ }\bibfield  {title} {\bibinfo {title} {General protocols for the efficient distillation of indistinguishable photons},\ }\href@noop {} {\bibfield  {journal} {\bibinfo  {journal} {Physical Review Applied}\ }\textbf {\bibinfo {volume} {23}},\ \bibinfo {pages} {034079} (\bibinfo {year} {2025})}\BibitemShut {NoStop}%
\bibitem [{\citenamefont {Flamini}\ \emph {et~al.}(2017)\citenamefont {Flamini}, \citenamefont {Spagnolo}, \citenamefont {Viggianiello}, \citenamefont {Crespi}, \citenamefont {Osellame},\ and\ \citenamefont {Sciarrino}}]{flamini2017benchmarking}%
  \BibitemOpen
  \bibfield  {author} {\bibinfo {author} {\bibfnamefont {F.}~\bibnamefont {Flamini}}, \bibinfo {author} {\bibfnamefont {N.}~\bibnamefont {Spagnolo}}, \bibinfo {author} {\bibfnamefont {N.}~\bibnamefont {Viggianiello}}, \bibinfo {author} {\bibfnamefont {A.}~\bibnamefont {Crespi}}, \bibinfo {author} {\bibfnamefont {R.}~\bibnamefont {Osellame}},\ and\ \bibinfo {author} {\bibfnamefont {F.}~\bibnamefont {Sciarrino}},\ }\bibfield  {title} {\bibinfo {title} {Benchmarking integrated linear-optical architectures for quantum information processing},\ }\href@noop {} {\bibfield  {journal} {\bibinfo  {journal} {Scientific Reports}\ }\textbf {\bibinfo {volume} {7}},\ \bibinfo {pages} {15133} (\bibinfo {year} {2017})}\BibitemShut {NoStop}%
\bibitem [{\citenamefont {Shaw}\ \emph {et~al.}(2023)\citenamefont {Shaw}, \citenamefont {Jones}, \citenamefont {Yard},\ and\ \citenamefont {Laing}}]{shaw2023errors}%
  \BibitemOpen
  \bibfield  {author} {\bibinfo {author} {\bibfnamefont {R.~D.}\ \bibnamefont {Shaw}}, \bibinfo {author} {\bibfnamefont {A.~E.}\ \bibnamefont {Jones}}, \bibinfo {author} {\bibfnamefont {P.}~\bibnamefont {Yard}},\ and\ \bibinfo {author} {\bibfnamefont {A.}~\bibnamefont {Laing}},\ }\bibfield  {title} {\bibinfo {title} {Errors in heralded circuits for linear optical entanglement generation},\ }\href@noop {} {\bibfield  {journal} {\bibinfo  {journal} {arXiv preprint arXiv:2305.08452}\ } (\bibinfo {year} {2023})}\BibitemShut {NoStop}%
\bibitem [{\citenamefont {Saied}\ \emph {et~al.}(2024)\citenamefont {Saied}, \citenamefont {Marshall}, \citenamefont {Anand}, \citenamefont {Grabbe},\ and\ \citenamefont {Rieffel}}]{saied2024advancing}%
  \BibitemOpen
  \bibfield  {author} {\bibinfo {author} {\bibfnamefont {J.}~\bibnamefont {Saied}}, \bibinfo {author} {\bibfnamefont {J.}~\bibnamefont {Marshall}}, \bibinfo {author} {\bibfnamefont {N.}~\bibnamefont {Anand}}, \bibinfo {author} {\bibfnamefont {S.}~\bibnamefont {Grabbe}},\ and\ \bibinfo {author} {\bibfnamefont {E.~G.}\ \bibnamefont {Rieffel}},\ }\bibfield  {title} {\bibinfo {title} {Advancing quantum networking: some tools and protocols for ideal and noisy photonic systems},\ }in\ \href@noop {} {\emph {\bibinfo {booktitle} {Quantum Computing, Communication, and Simulation IV}}},\ Vol.\ \bibinfo {volume} {12911}\ (\bibinfo {organization} {SPIE},\ \bibinfo {year} {2024})\ pp.\ \bibinfo {pages} {37--65}\BibitemShut {NoStop}%
\bibitem [{\citenamefont {Chanana}\ \emph {et~al.}(2022)\citenamefont {Chanana}, \citenamefont {Larocque}, \citenamefont {Moreira}, \citenamefont {Carolan}, \citenamefont {Guha}, \citenamefont {Melo}, \citenamefont {Anant}, \citenamefont {Song}, \citenamefont {Englund}, \citenamefont {Blumenthal} \emph {et~al.}}]{chanana2022ultra}%
  \BibitemOpen
  \bibfield  {author} {\bibinfo {author} {\bibfnamefont {A.}~\bibnamefont {Chanana}}, \bibinfo {author} {\bibfnamefont {H.}~\bibnamefont {Larocque}}, \bibinfo {author} {\bibfnamefont {R.}~\bibnamefont {Moreira}}, \bibinfo {author} {\bibfnamefont {J.}~\bibnamefont {Carolan}}, \bibinfo {author} {\bibfnamefont {B.}~\bibnamefont {Guha}}, \bibinfo {author} {\bibfnamefont {E.~G.}\ \bibnamefont {Melo}}, \bibinfo {author} {\bibfnamefont {V.}~\bibnamefont {Anant}}, \bibinfo {author} {\bibfnamefont {J.}~\bibnamefont {Song}}, \bibinfo {author} {\bibfnamefont {D.}~\bibnamefont {Englund}}, \bibinfo {author} {\bibfnamefont {D.~J.}\ \bibnamefont {Blumenthal}}, \emph {et~al.},\ }\bibfield  {title} {\bibinfo {title} {Ultra-low loss quantum photonic circuits integrated with single quantum emitters},\ }\href@noop {} {\bibfield  {journal} {\bibinfo  {journal} {Nature Communications}\ }\textbf {\bibinfo {volume} {13}},\ \bibinfo {pages} {7693} (\bibinfo {year} {2022})}\BibitemShut {NoStop}%
\bibitem [{\citenamefont {Taballione}\ \emph {et~al.}(2019)\citenamefont {Taballione}, \citenamefont {Wolterink}, \citenamefont {Lugani}, \citenamefont {Eckstein}, \citenamefont {Bell}, \citenamefont {Grootjans}, \citenamefont {Visscher}, \citenamefont {Geskus}, \citenamefont {Roeloffzen}, \citenamefont {Renema} \emph {et~al.}}]{taballione20198}%
  \BibitemOpen
  \bibfield  {author} {\bibinfo {author} {\bibfnamefont {C.}~\bibnamefont {Taballione}}, \bibinfo {author} {\bibfnamefont {T.~A.}\ \bibnamefont {Wolterink}}, \bibinfo {author} {\bibfnamefont {J.}~\bibnamefont {Lugani}}, \bibinfo {author} {\bibfnamefont {A.}~\bibnamefont {Eckstein}}, \bibinfo {author} {\bibfnamefont {B.~A.}\ \bibnamefont {Bell}}, \bibinfo {author} {\bibfnamefont {R.}~\bibnamefont {Grootjans}}, \bibinfo {author} {\bibfnamefont {I.}~\bibnamefont {Visscher}}, \bibinfo {author} {\bibfnamefont {D.}~\bibnamefont {Geskus}}, \bibinfo {author} {\bibfnamefont {C.~G.}\ \bibnamefont {Roeloffzen}}, \bibinfo {author} {\bibfnamefont {J.~J.}\ \bibnamefont {Renema}}, \emph {et~al.},\ }\bibfield  {title} {\bibinfo {title} {8$\times$ 8 reconfigurable quantum photonic processor based on silicon nitride waveguides},\ }\href@noop {} {\bibfield  {journal} {\bibinfo  {journal} {Optics express}\ }\textbf {\bibinfo {volume} {27}},\ \bibinfo {pages} {26842} (\bibinfo {year} {2019})}\BibitemShut {NoStop}%
\bibitem [{\citenamefont {Taballione}\ \emph {et~al.}(2023)\citenamefont {Taballione}, \citenamefont {Anguita}, \citenamefont {de~Goede}, \citenamefont {Venderbosch}, \citenamefont {Kassenberg}, \citenamefont {Snijders}, \citenamefont {Kannan}, \citenamefont {Vleeshouwers}, \citenamefont {Smith}, \citenamefont {Epping} \emph {et~al.}}]{taballione202320}%
  \BibitemOpen
  \bibfield  {author} {\bibinfo {author} {\bibfnamefont {C.}~\bibnamefont {Taballione}}, \bibinfo {author} {\bibfnamefont {M.~C.}\ \bibnamefont {Anguita}}, \bibinfo {author} {\bibfnamefont {M.}~\bibnamefont {de~Goede}}, \bibinfo {author} {\bibfnamefont {P.}~\bibnamefont {Venderbosch}}, \bibinfo {author} {\bibfnamefont {B.}~\bibnamefont {Kassenberg}}, \bibinfo {author} {\bibfnamefont {H.}~\bibnamefont {Snijders}}, \bibinfo {author} {\bibfnamefont {N.}~\bibnamefont {Kannan}}, \bibinfo {author} {\bibfnamefont {W.~L.}\ \bibnamefont {Vleeshouwers}}, \bibinfo {author} {\bibfnamefont {D.}~\bibnamefont {Smith}}, \bibinfo {author} {\bibfnamefont {J.~P.}\ \bibnamefont {Epping}}, \emph {et~al.},\ }\bibfield  {title} {\bibinfo {title} {20-mode universal quantum photonic processor},\ }\href@noop {} {\bibfield  {journal} {\bibinfo  {journal} {Quantum}\ }\textbf {\bibinfo {volume} {7}},\ \bibinfo {pages} {1071} (\bibinfo {year} {2023})}\BibitemShut {NoStop}%
\bibitem [{\citenamefont {Zhou}\ \emph {et~al.}(2022)\citenamefont {Zhou}, \citenamefont {Dong}, \citenamefont {Cheng}, \citenamefont {Dong}, \citenamefont {Huang}, \citenamefont {Shen}, \citenamefont {Zhang}, \citenamefont {Gu}, \citenamefont {Qian}, \citenamefont {Chen} \emph {et~al.}}]{zhou2022photonic}%
  \BibitemOpen
  \bibfield  {author} {\bibinfo {author} {\bibfnamefont {H.}~\bibnamefont {Zhou}}, \bibinfo {author} {\bibfnamefont {J.}~\bibnamefont {Dong}}, \bibinfo {author} {\bibfnamefont {J.}~\bibnamefont {Cheng}}, \bibinfo {author} {\bibfnamefont {W.}~\bibnamefont {Dong}}, \bibinfo {author} {\bibfnamefont {C.}~\bibnamefont {Huang}}, \bibinfo {author} {\bibfnamefont {Y.}~\bibnamefont {Shen}}, \bibinfo {author} {\bibfnamefont {Q.}~\bibnamefont {Zhang}}, \bibinfo {author} {\bibfnamefont {M.}~\bibnamefont {Gu}}, \bibinfo {author} {\bibfnamefont {C.}~\bibnamefont {Qian}}, \bibinfo {author} {\bibfnamefont {H.}~\bibnamefont {Chen}}, \emph {et~al.},\ }\bibfield  {title} {\bibinfo {title} {Photonic matrix multiplication lights up photonic accelerator and beyond},\ }\href@noop {} {\bibfield  {journal} {\bibinfo  {journal} {Light: Science \& Applications}\ }\textbf {\bibinfo {volume} {11}},\ \bibinfo {pages} {30} (\bibinfo {year} {2022})}\BibitemShut {NoStop}%
\bibitem [{\citenamefont {Erhard}\ \emph {et~al.}(2020)\citenamefont {Erhard}, \citenamefont {Krenn},\ and\ \citenamefont {Zeilinger}}]{erhard2020advances}%
  \BibitemOpen
  \bibfield  {author} {\bibinfo {author} {\bibfnamefont {M.}~\bibnamefont {Erhard}}, \bibinfo {author} {\bibfnamefont {M.}~\bibnamefont {Krenn}},\ and\ \bibinfo {author} {\bibfnamefont {A.}~\bibnamefont {Zeilinger}},\ }\bibfield  {title} {\bibinfo {title} {Advances in high-dimensional quantum entanglement},\ }\href@noop {} {\bibfield  {journal} {\bibinfo  {journal} {Nature Reviews Physics}\ }\textbf {\bibinfo {volume} {2}},\ \bibinfo {pages} {365} (\bibinfo {year} {2020})}\BibitemShut {NoStop}%
\bibitem [{\citenamefont {Bai}\ \emph {et~al.}(2023)\citenamefont {Bai}, \citenamefont {Xu}, \citenamefont {Tan}, \citenamefont {Sun}, \citenamefont {Li}, \citenamefont {Wu}, \citenamefont {Morandotti}, \citenamefont {Mitchell}, \citenamefont {Xu},\ and\ \citenamefont {Moss}}]{bai2023photonic}%
  \BibitemOpen
  \bibfield  {author} {\bibinfo {author} {\bibfnamefont {Y.}~\bibnamefont {Bai}}, \bibinfo {author} {\bibfnamefont {X.}~\bibnamefont {Xu}}, \bibinfo {author} {\bibfnamefont {M.}~\bibnamefont {Tan}}, \bibinfo {author} {\bibfnamefont {Y.}~\bibnamefont {Sun}}, \bibinfo {author} {\bibfnamefont {Y.}~\bibnamefont {Li}}, \bibinfo {author} {\bibfnamefont {J.}~\bibnamefont {Wu}}, \bibinfo {author} {\bibfnamefont {R.}~\bibnamefont {Morandotti}}, \bibinfo {author} {\bibfnamefont {A.}~\bibnamefont {Mitchell}}, \bibinfo {author} {\bibfnamefont {K.}~\bibnamefont {Xu}},\ and\ \bibinfo {author} {\bibfnamefont {D.~J.}\ \bibnamefont {Moss}},\ }\bibfield  {title} {\bibinfo {title} {Photonic multiplexing techniques for neuromorphic computing},\ }\href@noop {} {\bibfield  {journal} {\bibinfo  {journal} {Nanophotonics}\ }\textbf {\bibinfo {volume} {12}},\ \bibinfo {pages} {795} (\bibinfo {year} {2023})}\BibitemShut {NoStop}%
\bibitem [{\citenamefont {Bandyopadhyay}\ \emph {et~al.}(2024)\citenamefont {Bandyopadhyay}, \citenamefont {Sludds}, \citenamefont {Krastanov}, \citenamefont {Hamerly}, \citenamefont {Harris}, \citenamefont {Bunandar}, \citenamefont {Streshinsky}, \citenamefont {Hochberg},\ and\ \citenamefont {Englund}}]{bandyopadhyay2024single}%
  \BibitemOpen
  \bibfield  {author} {\bibinfo {author} {\bibfnamefont {S.}~\bibnamefont {Bandyopadhyay}}, \bibinfo {author} {\bibfnamefont {A.}~\bibnamefont {Sludds}}, \bibinfo {author} {\bibfnamefont {S.}~\bibnamefont {Krastanov}}, \bibinfo {author} {\bibfnamefont {R.}~\bibnamefont {Hamerly}}, \bibinfo {author} {\bibfnamefont {N.}~\bibnamefont {Harris}}, \bibinfo {author} {\bibfnamefont {D.}~\bibnamefont {Bunandar}}, \bibinfo {author} {\bibfnamefont {M.}~\bibnamefont {Streshinsky}}, \bibinfo {author} {\bibfnamefont {M.}~\bibnamefont {Hochberg}},\ and\ \bibinfo {author} {\bibfnamefont {D.}~\bibnamefont {Englund}},\ }\bibfield  {title} {\bibinfo {title} {Single-chip photonic deep neural network with forward-only training},\ }\href@noop {} {\bibfield  {journal} {\bibinfo  {journal} {Nature Photonics}\ }\textbf {\bibinfo {volume} {18}},\ \bibinfo {pages} {1335} (\bibinfo {year} {2024})}\BibitemShut {NoStop}%
\bibitem [{\citenamefont {Kitayama}\ \emph {et~al.}(2019)\citenamefont {Kitayama}, \citenamefont {Notomi}, \citenamefont {Naruse}, \citenamefont {Inoue}, \citenamefont {Kawakami},\ and\ \citenamefont {Uchida}}]{kitayama2019novel}%
  \BibitemOpen
  \bibfield  {author} {\bibinfo {author} {\bibfnamefont {K.-i.}\ \bibnamefont {Kitayama}}, \bibinfo {author} {\bibfnamefont {M.}~\bibnamefont {Notomi}}, \bibinfo {author} {\bibfnamefont {M.}~\bibnamefont {Naruse}}, \bibinfo {author} {\bibfnamefont {K.}~\bibnamefont {Inoue}}, \bibinfo {author} {\bibfnamefont {S.}~\bibnamefont {Kawakami}},\ and\ \bibinfo {author} {\bibfnamefont {A.}~\bibnamefont {Uchida}},\ }\bibfield  {title} {\bibinfo {title} {Novel frontier of photonics for data processing—photonic accelerator},\ }\href@noop {} {\bibfield  {journal} {\bibinfo  {journal} {Apl Photonics}\ }\textbf {\bibinfo {volume} {4}} (\bibinfo {year} {2019})}\BibitemShut {NoStop}%
\bibitem [{\citenamefont {Hilaire}\ \emph {et~al.}(2023)\citenamefont {Hilaire}, \citenamefont {Castor}, \citenamefont {Barnes}, \citenamefont {Economou},\ and\ \citenamefont {Grosshans}}]{hilaire2023linear}%
  \BibitemOpen
  \bibfield  {author} {\bibinfo {author} {\bibfnamefont {P.}~\bibnamefont {Hilaire}}, \bibinfo {author} {\bibfnamefont {Y.}~\bibnamefont {Castor}}, \bibinfo {author} {\bibfnamefont {E.}~\bibnamefont {Barnes}}, \bibinfo {author} {\bibfnamefont {S.~E.}\ \bibnamefont {Economou}},\ and\ \bibinfo {author} {\bibfnamefont {F.}~\bibnamefont {Grosshans}},\ }\bibfield  {title} {\bibinfo {title} {Linear optical logical bell state measurements with optimal loss-tolerance threshold},\ }\href@noop {} {\bibfield  {journal} {\bibinfo  {journal} {PRX Quantum}\ }\textbf {\bibinfo {volume} {4}},\ \bibinfo {pages} {040322} (\bibinfo {year} {2023})}\BibitemShut {NoStop}%
\bibitem [{\citenamefont {Giustina}\ \emph {et~al.}(2013)\citenamefont {Giustina}, \citenamefont {Mech}, \citenamefont {Ramelow}, \citenamefont {Wittmann}, \citenamefont {Kofler}, \citenamefont {Beyer}, \citenamefont {Lita}, \citenamefont {Calkins}, \citenamefont {Gerrits}, \citenamefont {Nam} \emph {et~al.}}]{giustina2013bell}%
  \BibitemOpen
  \bibfield  {author} {\bibinfo {author} {\bibfnamefont {M.}~\bibnamefont {Giustina}}, \bibinfo {author} {\bibfnamefont {A.}~\bibnamefont {Mech}}, \bibinfo {author} {\bibfnamefont {S.}~\bibnamefont {Ramelow}}, \bibinfo {author} {\bibfnamefont {B.}~\bibnamefont {Wittmann}}, \bibinfo {author} {\bibfnamefont {J.}~\bibnamefont {Kofler}}, \bibinfo {author} {\bibfnamefont {J.}~\bibnamefont {Beyer}}, \bibinfo {author} {\bibfnamefont {A.}~\bibnamefont {Lita}}, \bibinfo {author} {\bibfnamefont {B.}~\bibnamefont {Calkins}}, \bibinfo {author} {\bibfnamefont {T.}~\bibnamefont {Gerrits}}, \bibinfo {author} {\bibfnamefont {S.~W.}\ \bibnamefont {Nam}}, \emph {et~al.},\ }\bibfield  {title} {\bibinfo {title} {Bell violation using entangled photons without the fair-sampling assumption},\ }\href@noop {} {\bibfield  {journal} {\bibinfo  {journal} {Nature}\ }\textbf {\bibinfo {volume} {497}},\ \bibinfo {pages} {227} (\bibinfo {year} {2013})}\BibitemShut {NoStop}%
\bibitem [{\citenamefont {Eberhard}(1993)}]{eberhard1993background}%
  \BibitemOpen
  \bibfield  {author} {\bibinfo {author} {\bibfnamefont {P.~H.}\ \bibnamefont {Eberhard}},\ }\bibfield  {title} {\bibinfo {title} {Background level and counter efficiencies required for a loophole-free einstein-podolsky-rosen experiment},\ }\href@noop {} {\bibfield  {journal} {\bibinfo  {journal} {Physical Review A}\ }\textbf {\bibinfo {volume} {47}},\ \bibinfo {pages} {R747} (\bibinfo {year} {1993})}\BibitemShut {NoStop}%
\bibitem [{\citenamefont {Brunner}\ \emph {et~al.}(2007)\citenamefont {Brunner}, \citenamefont {Gisin}, \citenamefont {Scarani},\ and\ \citenamefont {Simon}}]{brunner2007detection}%
  \BibitemOpen
  \bibfield  {author} {\bibinfo {author} {\bibfnamefont {N.}~\bibnamefont {Brunner}}, \bibinfo {author} {\bibfnamefont {N.}~\bibnamefont {Gisin}}, \bibinfo {author} {\bibfnamefont {V.}~\bibnamefont {Scarani}},\ and\ \bibinfo {author} {\bibfnamefont {C.}~\bibnamefont {Simon}},\ }\bibfield  {title} {\bibinfo {title} {Detection loophole in asymmetric bell experiments},\ }\href@noop {} {\bibfield  {journal} {\bibinfo  {journal} {Physical review letters}\ }\textbf {\bibinfo {volume} {98}},\ \bibinfo {pages} {220403} (\bibinfo {year} {2007})}\BibitemShut {NoStop}%
\bibitem [{Note1()}]{Note1}%
  \BibitemOpen
  \bibinfo {note} {Here, $\rho $ may represent a state within a larger entangled system, such as in fusion gates \cite {bartolucci2023fusion}.}\BibitemShut {Stop}%
\bibitem [{\citenamefont {Tischler}\ \emph {et~al.}(2018)\citenamefont {Tischler}, \citenamefont {Rockstuhl},\ and\ \citenamefont {S{\l}owik}}]{tischler2018quantum}%
  \BibitemOpen
  \bibfield  {author} {\bibinfo {author} {\bibfnamefont {N.}~\bibnamefont {Tischler}}, \bibinfo {author} {\bibfnamefont {C.}~\bibnamefont {Rockstuhl}},\ and\ \bibinfo {author} {\bibfnamefont {K.}~\bibnamefont {S{\l}owik}},\ }\bibfield  {title} {\bibinfo {title} {Quantum optical realization of arbitrary linear transformations allowing for loss and gain},\ }\href@noop {} {\bibfield  {journal} {\bibinfo  {journal} {Physical Review X}\ }\textbf {\bibinfo {volume} {8}},\ \bibinfo {pages} {021017} (\bibinfo {year} {2018})}\BibitemShut {NoStop}%
\bibitem [{\citenamefont {Oszmaniec}\ and\ \citenamefont {Brod}(2018)}]{oszmaniec2018classical}%
  \BibitemOpen
  \bibfield  {author} {\bibinfo {author} {\bibfnamefont {M.}~\bibnamefont {Oszmaniec}}\ and\ \bibinfo {author} {\bibfnamefont {D.~J.}\ \bibnamefont {Brod}},\ }\bibfield  {title} {\bibinfo {title} {Classical simulation of photonic linear optics with lost particles},\ }\href@noop {} {\bibfield  {journal} {\bibinfo  {journal} {New Journal of Physics}\ }\textbf {\bibinfo {volume} {20}},\ \bibinfo {pages} {092002} (\bibinfo {year} {2018})}\BibitemShut {NoStop}%
\bibitem [{\citenamefont {Brod}\ and\ \citenamefont {Oszmaniec}(2020)}]{brod2020classical}%
  \BibitemOpen
  \bibfield  {author} {\bibinfo {author} {\bibfnamefont {D.~J.}\ \bibnamefont {Brod}}\ and\ \bibinfo {author} {\bibfnamefont {M.}~\bibnamefont {Oszmaniec}},\ }\bibfield  {title} {\bibinfo {title} {Classical simulation of linear optics subject to nonuniform losses},\ }\href@noop {} {\bibfield  {journal} {\bibinfo  {journal} {Quantum}\ }\textbf {\bibinfo {volume} {4}},\ \bibinfo {pages} {267} (\bibinfo {year} {2020})}\BibitemShut {NoStop}%
\bibitem [{\citenamefont {Melkozerov}\ \emph {et~al.}(2024)\citenamefont {Melkozerov}, \citenamefont {Avanesov}, \citenamefont {Dyakonov},\ and\ \citenamefont {Straupe}}]{melkozerov2024analysis}%
  \BibitemOpen
  \bibfield  {author} {\bibinfo {author} {\bibfnamefont {A.}~\bibnamefont {Melkozerov}}, \bibinfo {author} {\bibfnamefont {A.}~\bibnamefont {Avanesov}}, \bibinfo {author} {\bibfnamefont {I.}~\bibnamefont {Dyakonov}},\ and\ \bibinfo {author} {\bibfnamefont {S.}~\bibnamefont {Straupe}},\ }\bibfield  {title} {\bibinfo {title} {Analysis of optical loss thresholds in the fusion-based quantum computing architecture},\ }\href@noop {} {\bibfield  {journal} {\bibinfo  {journal} {APL Quantum}\ }\textbf {\bibinfo {volume} {1}} (\bibinfo {year} {2024})}\BibitemShut {NoStop}%
\bibitem [{\citenamefont {Bell}\ and\ \citenamefont {Walmsley}(2021)}]{bell2021further}%
  \BibitemOpen
  \bibfield  {author} {\bibinfo {author} {\bibfnamefont {B.~A.}\ \bibnamefont {Bell}}\ and\ \bibinfo {author} {\bibfnamefont {I.~A.}\ \bibnamefont {Walmsley}},\ }\bibfield  {title} {\bibinfo {title} {Further compactifying linear optical unitaries},\ }\href@noop {} {\bibfield  {journal} {\bibinfo  {journal} {Apl Photonics}\ }\textbf {\bibinfo {volume} {6}} (\bibinfo {year} {2021})}\BibitemShut {NoStop}%
\bibitem [{\citenamefont {Wang}\ \emph {et~al.}(2019)\citenamefont {Wang}, \citenamefont {Qin}, \citenamefont {Ding}, \citenamefont {Chen}, \citenamefont {Chen}, \citenamefont {You}, \citenamefont {He}, \citenamefont {Jiang}, \citenamefont {You}, \citenamefont {Wang} \emph {et~al.}}]{wang2019boson}%
  \BibitemOpen
  \bibfield  {author} {\bibinfo {author} {\bibfnamefont {H.}~\bibnamefont {Wang}}, \bibinfo {author} {\bibfnamefont {J.}~\bibnamefont {Qin}}, \bibinfo {author} {\bibfnamefont {X.}~\bibnamefont {Ding}}, \bibinfo {author} {\bibfnamefont {M.-C.}\ \bibnamefont {Chen}}, \bibinfo {author} {\bibfnamefont {S.}~\bibnamefont {Chen}}, \bibinfo {author} {\bibfnamefont {X.}~\bibnamefont {You}}, \bibinfo {author} {\bibfnamefont {Y.-M.}\ \bibnamefont {He}}, \bibinfo {author} {\bibfnamefont {X.}~\bibnamefont {Jiang}}, \bibinfo {author} {\bibfnamefont {L.}~\bibnamefont {You}}, \bibinfo {author} {\bibfnamefont {Z.}~\bibnamefont {Wang}}, \emph {et~al.},\ }\bibfield  {title} {\bibinfo {title} {Boson sampling with 20 input photons and a 60-mode interferometer in a 1 0 14-dimensional hilbert space},\ }\href@noop {} {\bibfield  {journal} {\bibinfo  {journal} {Physical review letters}\ }\textbf {\bibinfo {volume} {123}},\ \bibinfo {pages} {250503} (\bibinfo {year} {2019})}\BibitemShut {NoStop}%
\bibitem [{\citenamefont {Zhong}\ \emph {et~al.}(2020)\citenamefont {Zhong}, \citenamefont {Wang}, \citenamefont {Deng}, \citenamefont {Chen}, \citenamefont {Peng}, \citenamefont {Luo}, \citenamefont {Qin}, \citenamefont {Wu}, \citenamefont {Ding}, \citenamefont {Hu} \emph {et~al.}}]{zhong2020quantum}%
  \BibitemOpen
  \bibfield  {author} {\bibinfo {author} {\bibfnamefont {H.-S.}\ \bibnamefont {Zhong}}, \bibinfo {author} {\bibfnamefont {H.}~\bibnamefont {Wang}}, \bibinfo {author} {\bibfnamefont {Y.-H.}\ \bibnamefont {Deng}}, \bibinfo {author} {\bibfnamefont {M.-C.}\ \bibnamefont {Chen}}, \bibinfo {author} {\bibfnamefont {L.-C.}\ \bibnamefont {Peng}}, \bibinfo {author} {\bibfnamefont {Y.-H.}\ \bibnamefont {Luo}}, \bibinfo {author} {\bibfnamefont {J.}~\bibnamefont {Qin}}, \bibinfo {author} {\bibfnamefont {D.}~\bibnamefont {Wu}}, \bibinfo {author} {\bibfnamefont {X.}~\bibnamefont {Ding}}, \bibinfo {author} {\bibfnamefont {Y.}~\bibnamefont {Hu}}, \emph {et~al.},\ }\bibfield  {title} {\bibinfo {title} {Quantum computational advantage using photons},\ }\href@noop {} {\bibfield  {journal} {\bibinfo  {journal} {Science}\ }\textbf {\bibinfo {volume} {370}},\ \bibinfo {pages} {1460} (\bibinfo {year} {2020})}\BibitemShut {NoStop}%
\bibitem [{\citenamefont {Zhong}\ \emph {et~al.}(2021)\citenamefont {Zhong}, \citenamefont {Deng}, \citenamefont {Qin}, \citenamefont {Wang}, \citenamefont {Chen}, \citenamefont {Peng}, \citenamefont {Luo}, \citenamefont {Wu}, \citenamefont {Gong}, \citenamefont {Su} \emph {et~al.}}]{zhong2021phase}%
  \BibitemOpen
  \bibfield  {author} {\bibinfo {author} {\bibfnamefont {H.-S.}\ \bibnamefont {Zhong}}, \bibinfo {author} {\bibfnamefont {Y.-H.}\ \bibnamefont {Deng}}, \bibinfo {author} {\bibfnamefont {J.}~\bibnamefont {Qin}}, \bibinfo {author} {\bibfnamefont {H.}~\bibnamefont {Wang}}, \bibinfo {author} {\bibfnamefont {M.-C.}\ \bibnamefont {Chen}}, \bibinfo {author} {\bibfnamefont {L.-C.}\ \bibnamefont {Peng}}, \bibinfo {author} {\bibfnamefont {Y.-H.}\ \bibnamefont {Luo}}, \bibinfo {author} {\bibfnamefont {D.}~\bibnamefont {Wu}}, \bibinfo {author} {\bibfnamefont {S.-Q.}\ \bibnamefont {Gong}}, \bibinfo {author} {\bibfnamefont {H.}~\bibnamefont {Su}}, \emph {et~al.},\ }\bibfield  {title} {\bibinfo {title} {Phase-programmable gaussian boson sampling using stimulated squeezed light},\ }\href@noop {} {\bibfield  {journal} {\bibinfo  {journal} {Physical review letters}\ }\textbf {\bibinfo {volume} {127}},\ \bibinfo {pages} {180502} (\bibinfo {year} {2021})}\BibitemShut {NoStop}%
\bibitem [{\citenamefont {Madsen}\ \emph {et~al.}(2022)\citenamefont {Madsen}, \citenamefont {Laudenbach}, \citenamefont {Askarani}, \citenamefont {Rortais}, \citenamefont {Vincent}, \citenamefont {Bulmer}, \citenamefont {Miatto}, \citenamefont {Neuhaus}, \citenamefont {Helt}, \citenamefont {Collins} \emph {et~al.}}]{madsen2022quantum}%
  \BibitemOpen
  \bibfield  {author} {\bibinfo {author} {\bibfnamefont {L.~S.}\ \bibnamefont {Madsen}}, \bibinfo {author} {\bibfnamefont {F.}~\bibnamefont {Laudenbach}}, \bibinfo {author} {\bibfnamefont {M.~F.}\ \bibnamefont {Askarani}}, \bibinfo {author} {\bibfnamefont {F.}~\bibnamefont {Rortais}}, \bibinfo {author} {\bibfnamefont {T.}~\bibnamefont {Vincent}}, \bibinfo {author} {\bibfnamefont {J.~F.}\ \bibnamefont {Bulmer}}, \bibinfo {author} {\bibfnamefont {F.~M.}\ \bibnamefont {Miatto}}, \bibinfo {author} {\bibfnamefont {L.}~\bibnamefont {Neuhaus}}, \bibinfo {author} {\bibfnamefont {L.~G.}\ \bibnamefont {Helt}}, \bibinfo {author} {\bibfnamefont {M.~J.}\ \bibnamefont {Collins}}, \emph {et~al.},\ }\bibfield  {title} {\bibinfo {title} {Quantum computational advantage with a programmable photonic processor},\ }\href@noop {} {\bibfield  {journal} {\bibinfo  {journal} {Nature}\ }\textbf {\bibinfo {volume} {606}},\ \bibinfo {pages} {75} (\bibinfo {year} {2022})}\BibitemShut {NoStop}%
\bibitem [{\citenamefont {Deng}\ \emph {et~al.}(2023)\citenamefont {Deng}, \citenamefont {Gu}, \citenamefont {Liu}, \citenamefont {Gong}, \citenamefont {Su}, \citenamefont {Zhang}, \citenamefont {Tang}, \citenamefont {Jia}, \citenamefont {Xu}, \citenamefont {Chen} \emph {et~al.}}]{deng2023gaussian}%
  \BibitemOpen
  \bibfield  {author} {\bibinfo {author} {\bibfnamefont {Y.-H.}\ \bibnamefont {Deng}}, \bibinfo {author} {\bibfnamefont {Y.-C.}\ \bibnamefont {Gu}}, \bibinfo {author} {\bibfnamefont {H.-L.}\ \bibnamefont {Liu}}, \bibinfo {author} {\bibfnamefont {S.-Q.}\ \bibnamefont {Gong}}, \bibinfo {author} {\bibfnamefont {H.}~\bibnamefont {Su}}, \bibinfo {author} {\bibfnamefont {Z.-J.}\ \bibnamefont {Zhang}}, \bibinfo {author} {\bibfnamefont {H.-Y.}\ \bibnamefont {Tang}}, \bibinfo {author} {\bibfnamefont {M.-H.}\ \bibnamefont {Jia}}, \bibinfo {author} {\bibfnamefont {J.-M.}\ \bibnamefont {Xu}}, \bibinfo {author} {\bibfnamefont {M.-C.}\ \bibnamefont {Chen}}, \emph {et~al.},\ }\bibfield  {title} {\bibinfo {title} {Gaussian boson sampling with pseudo-photon-number-resolving detectors and quantum computational advantage},\ }\href@noop {} {\bibfield  {journal} {\bibinfo  {journal} {Physical review letters}\ }\textbf {\bibinfo {volume} {131}},\ \bibinfo {pages} {150601} (\bibinfo {year} {2023})}\BibitemShut {NoStop}%
\bibitem [{\citenamefont {Mezzadri}(2006)}]{mezzadri2006generate}%
  \BibitemOpen
  \bibfield  {author} {\bibinfo {author} {\bibfnamefont {F.}~\bibnamefont {Mezzadri}},\ }\bibfield  {title} {\bibinfo {title} {How to generate random matrices from the classical compact groups},\ }\href@noop {} {\bibfield  {journal} {\bibinfo  {journal} {arXiv preprint math-ph/0609050}\ } (\bibinfo {year} {2006})}\BibitemShut {NoStop}%
\bibitem [{\citenamefont {Tichy}\ \emph {et~al.}(2010)\citenamefont {Tichy}, \citenamefont {Tiersch}, \citenamefont {de~Melo}, \citenamefont {Mintert},\ and\ \citenamefont {Buchleitner}}]{tichy2010zero}%
  \BibitemOpen
  \bibfield  {author} {\bibinfo {author} {\bibfnamefont {M.~C.}\ \bibnamefont {Tichy}}, \bibinfo {author} {\bibfnamefont {M.}~\bibnamefont {Tiersch}}, \bibinfo {author} {\bibfnamefont {F.}~\bibnamefont {de~Melo}}, \bibinfo {author} {\bibfnamefont {F.}~\bibnamefont {Mintert}},\ and\ \bibinfo {author} {\bibfnamefont {A.}~\bibnamefont {Buchleitner}},\ }\bibfield  {title} {\bibinfo {title} {Zero-transmission law for multiport beam splitters},\ }\href@noop {} {\bibfield  {journal} {\bibinfo  {journal} {Physical review letters}\ }\textbf {\bibinfo {volume} {104}},\ \bibinfo {pages} {220405} (\bibinfo {year} {2010})}\BibitemShut {NoStop}%
\bibitem [{\citenamefont {Varnava}\ \emph {et~al.}(2008)\citenamefont {Varnava}, \citenamefont {Browne},\ and\ \citenamefont {Rudolph}}]{varnava2008good}%
  \BibitemOpen
  \bibfield  {author} {\bibinfo {author} {\bibfnamefont {M.}~\bibnamefont {Varnava}}, \bibinfo {author} {\bibfnamefont {D.~E.}\ \bibnamefont {Browne}},\ and\ \bibinfo {author} {\bibfnamefont {T.}~\bibnamefont {Rudolph}},\ }\bibfield  {title} {\bibinfo {title} {How good must single photon sources and detectors be for efficient linear optical quantum computation?},\ }\href@noop {} {\bibfield  {journal} {\bibinfo  {journal} {Physical review letters}\ }\textbf {\bibinfo {volume} {100}},\ \bibinfo {pages} {060502} (\bibinfo {year} {2008})}\BibitemShut {NoStop}%
\bibitem [{\citenamefont {Bartolucci}\ \emph {et~al.}(2023)\citenamefont {Bartolucci}, \citenamefont {Birchall}, \citenamefont {Bombin}, \citenamefont {Cable}, \citenamefont {Dawson}, \citenamefont {Gimeno-Segovia}, \citenamefont {Johnston}, \citenamefont {Kieling}, \citenamefont {Nickerson}, \citenamefont {Pant} \emph {et~al.}}]{bartolucci2023fusion}%
  \BibitemOpen
  \bibfield  {author} {\bibinfo {author} {\bibfnamefont {S.}~\bibnamefont {Bartolucci}}, \bibinfo {author} {\bibfnamefont {P.}~\bibnamefont {Birchall}}, \bibinfo {author} {\bibfnamefont {H.}~\bibnamefont {Bombin}}, \bibinfo {author} {\bibfnamefont {H.}~\bibnamefont {Cable}}, \bibinfo {author} {\bibfnamefont {C.}~\bibnamefont {Dawson}}, \bibinfo {author} {\bibfnamefont {M.}~\bibnamefont {Gimeno-Segovia}}, \bibinfo {author} {\bibfnamefont {E.}~\bibnamefont {Johnston}}, \bibinfo {author} {\bibfnamefont {K.}~\bibnamefont {Kieling}}, \bibinfo {author} {\bibfnamefont {N.}~\bibnamefont {Nickerson}}, \bibinfo {author} {\bibfnamefont {M.}~\bibnamefont {Pant}}, \emph {et~al.},\ }\bibfield  {title} {\bibinfo {title} {Fusion-based quantum computation},\ }\href@noop {} {\bibfield  {journal} {\bibinfo  {journal} {Nature Communications}\ }\textbf {\bibinfo {volume} {14}},\ \bibinfo {pages} {912} (\bibinfo {year} {2023})}\BibitemShut {NoStop}%
\bibitem [{\citenamefont {Chen}\ \emph {et~al.}(2024)\citenamefont {Chen}, \citenamefont {Peng}, \citenamefont {Guo}, \citenamefont {Gu}, \citenamefont {Ding}, \citenamefont {Liu}, \citenamefont {Zhao}, \citenamefont {You}, \citenamefont {Qin}, \citenamefont {Wang} \emph {et~al.}}]{chen2024heralded}%
  \BibitemOpen
  \bibfield  {author} {\bibinfo {author} {\bibfnamefont {S.}~\bibnamefont {Chen}}, \bibinfo {author} {\bibfnamefont {L.-C.}\ \bibnamefont {Peng}}, \bibinfo {author} {\bibfnamefont {Y.-P.}\ \bibnamefont {Guo}}, \bibinfo {author} {\bibfnamefont {X.-M.}\ \bibnamefont {Gu}}, \bibinfo {author} {\bibfnamefont {X.}~\bibnamefont {Ding}}, \bibinfo {author} {\bibfnamefont {R.-Z.}\ \bibnamefont {Liu}}, \bibinfo {author} {\bibfnamefont {J.-Y.}\ \bibnamefont {Zhao}}, \bibinfo {author} {\bibfnamefont {X.}~\bibnamefont {You}}, \bibinfo {author} {\bibfnamefont {J.}~\bibnamefont {Qin}}, \bibinfo {author} {\bibfnamefont {Y.-F.}\ \bibnamefont {Wang}}, \emph {et~al.},\ }\bibfield  {title} {\bibinfo {title} {Heralded three-photon entanglement from a single-photon source on a photonic chip},\ }\href@noop {} {\bibfield  {journal} {\bibinfo  {journal} {Physical Review Letters}\ }\textbf {\bibinfo {volume} {132}},\ \bibinfo {pages} {130603} (\bibinfo {year} {2024})}\BibitemShut {NoStop}%
\bibitem [{\citenamefont {Migdall}\ \emph {et~al.}(2002)\citenamefont {Migdall}, \citenamefont {Branning},\ and\ \citenamefont {Castelletto}}]{migdall2002tailoring}%
  \BibitemOpen
  \bibfield  {author} {\bibinfo {author} {\bibfnamefont {A.~L.}\ \bibnamefont {Migdall}}, \bibinfo {author} {\bibfnamefont {D.}~\bibnamefont {Branning}},\ and\ \bibinfo {author} {\bibfnamefont {S.}~\bibnamefont {Castelletto}},\ }\bibfield  {title} {\bibinfo {title} {Tailoring single-photon and multiphoton probabilities of a single-photon on-demand source},\ }\href@noop {} {\bibfield  {journal} {\bibinfo  {journal} {Physical Review A}\ }\textbf {\bibinfo {volume} {66}},\ \bibinfo {pages} {053805} (\bibinfo {year} {2002})}\BibitemShut {NoStop}%
\bibitem [{\citenamefont {Taballione}\ \emph {et~al.}(2021)\citenamefont {Taballione}, \citenamefont {van~der Meer}, \citenamefont {Snijders}, \citenamefont {Hooijschuur}, \citenamefont {Epping}, \citenamefont {de~Goede}, \citenamefont {Kassenberg}, \citenamefont {Venderbosch}, \citenamefont {Toebes}, \citenamefont {van~den Vlekkert} \emph {et~al.}}]{taballione2021universal}%
  \BibitemOpen
  \bibfield  {author} {\bibinfo {author} {\bibfnamefont {C.}~\bibnamefont {Taballione}}, \bibinfo {author} {\bibfnamefont {R.}~\bibnamefont {van~der Meer}}, \bibinfo {author} {\bibfnamefont {H.~J.}\ \bibnamefont {Snijders}}, \bibinfo {author} {\bibfnamefont {P.}~\bibnamefont {Hooijschuur}}, \bibinfo {author} {\bibfnamefont {J.~P.}\ \bibnamefont {Epping}}, \bibinfo {author} {\bibfnamefont {M.}~\bibnamefont {de~Goede}}, \bibinfo {author} {\bibfnamefont {B.}~\bibnamefont {Kassenberg}}, \bibinfo {author} {\bibfnamefont {P.}~\bibnamefont {Venderbosch}}, \bibinfo {author} {\bibfnamefont {C.}~\bibnamefont {Toebes}}, \bibinfo {author} {\bibfnamefont {H.}~\bibnamefont {van~den Vlekkert}}, \emph {et~al.},\ }\bibfield  {title} {\bibinfo {title} {A universal fully reconfigurable 12-mode quantum photonic processor},\ }\href@noop {} {\bibfield  {journal} {\bibinfo  {journal} {Materials for Quantum Technology}\ }\textbf {\bibinfo {volume} {1}},\ \bibinfo {pages} {035002} (\bibinfo {year} {2021})}\BibitemShut {NoStop}%
\end{thebibliography}%

\end{document}